\documentclass[12pt]{article} 
\usepackage{amssymb,amsmath,amsfonts,theorem,subfigure} 
\usepackage{graphicx}
\usepackage{epic,curves}
\usepackage{multirow}
\usepackage{pstricks}
\usepackage{array}
 
\newcommand{\lw}{\psset{linewidth=0.5pt}}
\newcommand{\unlw}{\psset{linewidth=1pt}}
\newrgbcolor{myc}{1 0 0}
\newrgbcolor{myc2}{0 0 1}
\newrgbcolor{myc3}{1 0 1}

\theorembodyfont{\itshape} 
\theoremheaderfont{\scshape}
\theoremstyle{plain}  
\newtheorem{Lemme}{Lemma}[section]

\newtheorem{Proposition}[Lemme]{Proposition}

\newtheorem{Conjecture}[Lemme]{Conjecture}

\newtheorem{Definition}{Definition}[section]

\title{{\huge A Proof of Selection Rules \\for Critical Dense Polymers}}
\author{
{\Large Alexi Morin-Duchesne}\footnote{\ttfamily alexi.morin-duchesne{\char'100}umontreal.ca},\\
\it D\'epartement de physique\\ 
\it Universit\'e de Montr\'eal, C.P.\ 6128, succ.\ centre-ville, Montr\'eal\\ 
\it Qu\'ebec, Canada, H3C 3J7\\[10pt]}

\begin{document} 

\maketitle

%
%
 
\begin{abstract}

Among the lattice loop models defined by Pearce, Rasmussen and Zuber (2006), the model corresponding to critical dense polymers ($\beta = 0$) is the only one for which an inversion relation for the transfer matrix $D_N(u)$ was found by Pearce and Rasmussen (2007). From this result, they identified the set of possible eigenvalues for $D_N(u)$ and gave a conjecture for the degeneracies of its relevant eigenvalues in the link representation, in the sector with $d$ defects. In this paper, we set out to prove this conjecture, using the homomorphism of the $TL_N (\beta)$ algebra between the loop model link representation and that of the XXZ model for $\beta = -(q+q^{-1})$. 
 \\


\end{abstract}

%
%

\tableofcontents

%
\section{Introduction} \label{sec:intro}
%

This paper proves a recent conjecture by Pearce and Rasmussen \cite{PR} for the model of critical dense polymers on the strip, by using the relation between this model and the Heisenberg spin model. The Heisenberg model (or XXZ model) is a long studied family of Hamiltonians of $N$ interacting spins on a chain. The models depend upon a spectral parameter $q$, which controls the $z$ interaction between neighboring spins. The Hamiltonian $H_{XXZ}$ acts on $(\mathbb C^2)^{\otimes N}$ (every spin is $\frac12$) and commutes with $S^z$. The spectrum of the XXX Hamiltonian ($q=1$) for the periodic chain was computed by Bethe \cite{Bethe} long ago and his method, the Bethe ansatz, has since allowed for solutions of the more general XXZ problem on various geometries (\cite{Gaudin}, \cite{Korepin}). In this paper, we focus on the case where the chain is finite and the Hamiltonian has very particular boundary terms for which the model is invariant under $U_q(sl_2)$ \cite{PasquierSaleur}. This symmetry will play an important role. We shall be particularly interested in the case $q=i$, for which the $z$ coupling in the Hamiltonian is absent (known as the XX-model). Though the Bethe ansatz solution is known, the spectrum of this Hamiltonian can be found using the simpler technique of Jordan-Wigner transformation \cite{JordanWigner}.

The loop models introduced in \cite{PRZ} are two dimensional lattice models on the strip that obey Yang-Baxter relations and are, in this sense, integrable. The transfer matrix $D_N(u)$ and Hamiltonian $\mathcal H_N$ of the model are elements of the Temperley-Lieb algebra $TL_N(\beta)$ and depend on one free parameter, the fugacity $\beta$ of the loops. The action of $TL_N(\beta)$ connectivities on link states (i.e. \hspace{-0.2cm} on $V_N$, the space they generate) defines a representation $\rho$ of $TL_N(\beta)$. For a given connectivity $c$, the matrix $\rho(c)$ is upper triangular (the number of defects, $d$, is a non increasing quantity) and its spectrum $\rho(c)$ is the union of the spectrums of the diagonal blocks, indexed by $d$, the number of defects. Moreover, the partition functions of Potts models and Fortuin-Kasteleyn models can be computed from the eigenvalues of $\rho(D_N(u))$ of the loop models for specific values of $\beta$ (\cite{Jacobsen}, \cite{JacobsenSaleur}, \cite{AMDSA}).

These models have attracted much interest because the $\rho$ representation of the Hamiltonian and transfer matrix exhibit non trivial Jordan cells (\cite{PRZ}, \cite{AMDSA}, \cite{Vasseur}). The corresponding representations of the Virasoro algebra should then be indecomposable and the underlying conformal field theory, logarithmic \cite{PRZ}. On the finite lattice, the diagonal blocks $\rho(D_N)|_d$ have been conjectured to be diagonalizable for $\beta \in [-2,2]$ for all $d$. Non trivial Jordan cells do occur, but they tie eigenvalues belonging to sectors with different numbers of defects. This structure appears for specific values of the fugacity $\beta = - (q + q^{-1})$ when $q$ is a root of unity. 

The case $\beta = 0$ is somewhat special, as an inversion relation for the transfer matrix was found \cite{PR}: $D_N(u)D_N(u+\frac{\pi}2)$ is a scalar multiple of the identity. From this, one can identify the set of all possible eigenvalues, and the degeneracies of each of these in a given sector $d$ was conjectured by Pearce and Rasmussen through selection rules \cite{PR}.

The two models introduced previously are known to be related (for example in \cite{deGier}, \cite{Nichols} and \cite{Vasseur}). Namely, there exists a $TL_N$-homomorphism $i_N^d$ of  from $V_N^d$ to $(\mathbb C^2)^{\otimes N}|_{S^z=d/2}$ (the restriction of $(\mathbb C^2)^{\otimes N}$ to spin configurations with $n = (N-d)/2$ down spins). The Heisenberg Hamiltonians can be expressed in terms of some matrices $e_i$s that act on $i_N^d(V_N^d)$ in the same way the Temperley-Lieb generators $U_i$ act on $V_N^d$ for $\beta = -(q +q^{-1})$ (except that the number of defects is conserved). For any $q$ and $\beta$ satisfying this relation and any $c \in TL_N(\beta)$, the spectrum of $\rho(c)$ can be found in the spectrum of $X(c)$, the representation of $c$ in the XXZ model. We will use the homomorphism to compute the degeneracies of $\rho(\mathcal H_N)$ and show they are those predicted by Pearce and Rasmussen \cite{PR}.

The outline of the paper is as follows. In section \ref{sec:intro2}, we review the definition of Temperley-Lieb algebra and of the transfer matrix for critical dense polymers. We recall the selection rules conjectured in \cite{PR} and translate these in terms of eigenvalue degeneracies of the Hamiltonian. In section \ref{sec:XXZ}, we perform the Jordan-Wigner transformation on the $XX$ Hamiltonian and write it in terms of creation and annihilation operators. For $N$ odd, we find $H_{XX}$ to be diagonalizable, but not for $N$ even, for which we provide its Jordan form (some technical details for $N$ even are given in appendix \ref{app:a}). The Hamiltonian $H_{XXZ}$ is invariant under $U_{q}(sl_2)$ and, in section \ref{sec:uqsl2}, we write down the generators of the $U_{q=i}(sl_2)$ algebra in terms of the creation and annihilation operators of section \ref{sec:XXZ}. In section \ref{sec:homo}, we explicit the homomorphism $i_N^d$ between $V_N^d$ and $\mathbb (C^2)^{\otimes N}|_{S^z=d/2}$, the vector space generated by spin configurations with $d$ down spins. We show that $i_N^d$ sends link states to $(\mathbb C^2)^{\otimes N}|_{S^z=d/2} \cap \ker(S^+)$. Because this homomorphism is injective, one can find the spectrum of any element of $TL_N(\beta)$ by looking at its representation in the Heisenberg problem. This is the goal of section $\ref{sec:reduction}$: we find a set of eigenvectors  that complement those in $i_N^d(V_N^d)$ and prove in appendix \ref{app:b} that these states are indeed independent. From this we can identify degeneracies in the XX Hamiltonian of eigenvectors $\in \ker(S^+)$ and show they reproduce the spectrum given by the selection rules in section \ref{sec:intro2}. 

%
\section{Critical dense polymers and selection rules} \label{sec:intro2}
%
\subsection{The algebra $TL_N(\beta)$ and the double-row matrix}

We start this section by recalling known definitions and results for the Temperley-Lieb algebra and its transfer matrices. The Temperley-Lieb algebra $TL_N(\beta)$ is a finite algebra, with generators $id, U_1, ..., U_{N-1}$ satisfying the relations
\begin{alignat}{3}
U_i^2&=\beta U_i,&\qquad & \nonumber\\
U_iU_j&=U_jU_i, &&\text{\rm for }|i-j|>1,\label{eq:TLdef}\\
U_iU_{i\pm 1}U_i&=U_i,&& \text{\rm  when $i, i\pm 1 \in \{1,2, \dots, N-1\}$.} \nonumber
\end{alignat}

The algebra $TL_N(\beta)$ is sometimes referred to as a connectivity algebra. A connectivity is a diagram made of a rectangular box with $N$ marked points on the top segment and as many marked points on the bottom. Inside the box, the $2N$ points are connected pairwise by non intersecting curves. To the generator $U_i$, we associate the connectivity
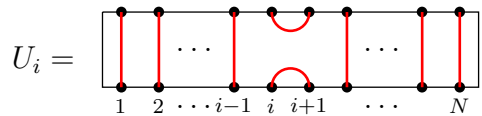
\begin{equation*} U_i =
\begin{pspicture}(0,-0.25)(5.5,0.5)
\psset{unit=0.5}
\psline[linewidth=0.5pt]{-}(0.5,1)(10.5,1)(10.5,-1)(0.5,-1)(0.5,1)
\psset{linewidth=1pt}
\psdots(1,1)(2,1)(4,1)(5,1)(6,1)(7,1)(9,1)(10,1)
(1,-1)(2,-1)(4,-1)(5,-1)(6,-1)(7,-1)(9,-1)(10,-1)
\psset{linecolor=myc}
\psline{-}(1,1)(1,-1)
\psline{-}(2,1)(2,-1)
\psline{-}(4,1)(4,-1)
\psline{-}(7,1)(7,-1)
\psline{-}(9,1)(9,-1)
\psline{-}(10,1)(10,-1)
\rput(3,0){$\dots$}
\rput(8,0){$\dots$}
\psarc(5.5,1){0.5}{180}{360}
\psarc(5.5,-1){0.5}{0}{180}
\rput(1,-1.5){$_1$}
\rput(2,-1.5){$_2$}
\rput(3,-1.5){\dots}
\rput(4,-1.5){$_{i-1}$}
\rput(5,-1.5){$_{i}$}
\rput(6,-1.5){$_{i+1}$}
\rput(8,-1.5){\dots}
\rput(10,-1.5){$_N$}
\end{pspicture}
\end{equation*}
\vspace{-0.2cm}

Diagrammatically, the product  $U_i U_j$ amounts to gluing the diagram of $U_j$ over the diagram of $U_i$. The resulting connectivity is obtained by reading the connections between the top and bottom marked points. With this identification, the first equation of (\ref{eq:TLdef}) is
\begin{equation*} U_i^2 =
\begin{pspicture}(0,-0.625)(5.5,0.5)
\psset{unit=0.5}
\psline[linewidth=0.5pt]{-}(0.5,1)(10.5,1)(10.5,-1)(0.5,-1)(0.5,1)
\psline[linewidth=0.5pt]{-}(0.5,-3)(10.5,-3)(10.5,-1)(0.5,-1)(0.5,-3)
\psset{linewidth=1pt}
\psdots(1,1)(2,1)(4,1)(5,1)(6,1)(7,1)(9,1)(10,1)
(1,-1)(2,-1)(4,-1)(5,-1)(6,-1)(7,-1)(9,-1)(10,-1)
(1,-3)(2,-3)(4,-3)(5,-3)(6,-3)(7,-3)(9,-3)(10,-3)
\psset{linecolor=myc}
\psline{-}(1,1)(1,-3)
\psline{-}(2,1)(2,-3)
\psline{-}(4,1)(4,-3)
\psline{-}(7,1)(7,-3)
\psline{-}(9,1)(9,-3)
\psline{-}(10,1)(10,-3)
\rput(3,0){$\dots$}
\rput(8,0){$\dots$}
\rput(3,-2){$\dots$}
\rput(8,-2){$\dots$}
\psarc(5.5,1){0.5}{180}{360}
\psarc(5.5,-1){0.5}{0}{360}
\psarc(5.5,-3){0.5}{0}{180}
\end{pspicture} = \beta U_i,
\end{equation*}
\vspace{0.3cm}
\\
 so that the free parameter $\beta$ is the weight given to loops closed in the process. The other two equations in (\ref{eq:TLdef}) have similar interpretations. Any connectivity can be obtained by a product of the generators, and the product of any two connectivities $c_1$ and $c_2$ in $TL_N(\beta)$ is given by the same concatenation rule. The algebra $TL_N(\beta)$ is the algebra of connectivities endowed with the product just described and is of dimension $\frac{1}{n+1}\left(\begin{smallmatrix} 2n \\ n \end{smallmatrix}\right)$.
 
 A useful representation is the representation $\rho$ on {\it link states} (or {\it link patterns}). A link pattern is a set of $N$ marked points on a horizontal segment. The points are connected pairwise, or to infinity, by non intersecting curves that lay above the segment. Points connected to infinity are called {\it defects}. The set of link states of length $N$ with $d$ defects is denoted $B_N^d$ and their linear span by $V_N^d$, with $\dim(V_N^d) = \left( \begin{smallmatrix} N \\ (N-d)/2\end{smallmatrix}\right) -  \left( \begin{smallmatrix} N \\ (N-d-2)/2\end{smallmatrix}\right)$. The set of all link states of size $N$ is noted $B_N$ (and $V_N$ the corresponding vector space). Let $v \in B_N$ and $c$ a connectivity. The product $c v$ is obtained by connecting the marked points of $v$ to the top marked points of $c$, by reading the resulting link pattern given by the new connections at the bottom of $c$, and by adding a multiplicative factor of $\beta$ for each closed loop. Here is an example:
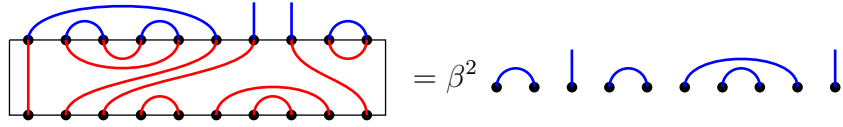
\vspace{-0.1cm}
\begin{equation*}
\begin{pspicture}(0,-0.125)(5.5,0.5)
\psset{unit=0.5}
\psline[linewidth=0.5pt]{-}(0.5,1)(10.5,1)(10.5,-1)(0.5,-1)(0.5,1)
\psset{linewidth=1pt}
\psdots(1,1)(2,1)(3,1)(4,1)(5,1)(6,1)(7,1)(8,1)(9,1)(10,1)
(1,-1)(2,-1)(3,-1)(4,-1)(5,-1)(6,-1)(7,-1)(8,-1)(9,-1)(10,-1)
\psset{linecolor=myc}
\psline{-}(1,1)(1,-1)
\psarc(3.5,1){0.5}{180}{360}
\psarc(9.5,1){0.5}{180}{360}
\psarc(4.5,-1){0.5}{0}{180}
\psarc(7.5,-1){0.5}{0}{180}
\psbezier{-}(8,1)(8,0)(10,0)(10,-1)
\psbezier{-}(9,-1)(9,0)(6,0)(6,-1)
\psbezier{-}(7,1)(7,0)(3,0)(3,-1)
\psbezier{-}(6,1)(6,0)(2,0)(2,-1)
\psbezier{-}(2,1)(2,0)(5,0)(5,1)
\psset{linecolor=myc2}
\psline{-}(7,1)(7,2)
\psline{-}(8,1)(8,2)
\psarc{-}(2.5,1){0.5}{0}{180}
\psarc{-}(4.5,1){0.5}{0}{180}
\psarc{-}(9.5,1){0.5}{0}{180}
\psbezier{-}(1,1)(1,2.2)(6,2.2)(6,1)
\end{pspicture} = \beta^2
\begin{pspicture}(0.25,-0.5)(5.5,0.5)
\psset{unit=0.5}
\psset{linewidth=1pt}
\psdots
(1,-1)(2,-1)(3,-1)(4,-1)(5,-1)(6,-1)(7,-1)(8,-1)(9,-1)(10,-1)
\psset{linecolor=myc2}
\psline{-}(10,-1)(10,0)
\psline{-}(3,-1)(3,0)
\psarc(4.5,-1){0.5}{0}{180}
\psarc(7.5,-1){0.5}{0}{180}
\psarc{-}(1.5,-1){0.5}{0}{180}
\psbezier{-}(9,-1)(9,0)(6,0)(6,-1)
\end{pspicture}
\end{equation*}
 
The matrix representing $c$ in the link representation is denoted $\rho(c)$. It is of size $\dim(V_N)$ and obtained by acting on $c$ with all the link patterns of $B_N$.
  We introduce the double-row matrix $D_N(u)$ as an element of  $TL_N(\beta = 0)$. It is defined diagrammatically by
\vspace{-0.4cm}
\begin{equation*}
\psset{unit=1.1}
\psset{linewidth=1pt}
\begin{pspicture}(-2,0)(0,2.2)
\rput(-2.0,1){$D_N(u) = \frac{\displaystyle{1}}{\displaystyle{\sin 2u}}$}
\end{pspicture}
\overbrace{
\begin{pspicture}(-0,0)(5,2.2)
\psdots(0.5,0)(1.5,0)(4.5,0)
\psdots(0.5,2)(1.5,2)(4.5,2)
\lw
\psline{-}(0,0)(1,0)(1,1)(0,1)(0,0)\psarc[linewidth=0.5pt]{-}(0,0){0.25}{0}{90}\rput(0.5,0.5){$u$}
\psline{-}(1,0)(2,0)(2,1)(1,1)(1,0)\psarc[linewidth=0.5pt]{-}(1,0){0.25}{0}{90}\rput(1.5,0.5){$u$}
\psline{-}(4,0)(5,0)(5,1)(4,1)(4,0)\psarc[linewidth=0.5pt]{-}(4,0){0.25}{0}{90}\rput(4.5,0.5){$u$}
\psline{-}(0,1)(1,1)(1,2)(0,2)(0,1)\psarc[linewidth=0.5pt]{-}(0,1){0.25}{0}{90}\rput(0.5,1.5){$\frac\pi2-u$}
\psline{-}(1,1)(2,1)(2,2)(1,2)(1,1)\psarc[linewidth=0.5pt]{-}(1,1){0.25}{0}{90}\rput(1.5,1.5){$\frac\pi2-u$}
\psline{-}(4,1)(5,1)(5,2)(4,2)(4,1)\psarc[linewidth=0.5pt]{-}(4,1){0.25}{0}{90}\rput(4.5,1.5){$\frac\pi2-u$}
\psline{-}(2,0)(2.5,0)\psline[linestyle=dashed,dash=2pt 2pt]{-}(2.5,0)(3.5,0)\psline{-}(3.5,0)(4,0)
\psline{-}(2,1)(2.5,1)\psline[linestyle=dashed,dash=2pt 2pt]{-}(2.5,1)(3.5,1)\psline{-}(3.5,1)(4,1)
\psline{-}(2,2)(2.5,2)\psline[linestyle=dashed,dash=2pt 2pt]{-}(2.5,2)(3.5,2)\psline{-}(3.5,2)(4,2)
\psset{linecolor=myc}\unlw
\psarc{-}(0,1){0.5}{90}{270}
\psarc{-}(5,1){0.5}{270}{450}
\end{pspicture}}^N
\end{equation*}
where each box is given by
\begin{equation*}
\psset{linewidth=1pt}
\begin{pspicture}(-0.5,-0.1)(0.5,0.5)\lw
\psline{-}(-0.5,-0.5)(0.5,-0.5)(0.5,0.5)(-0.5,0.5)(-0.5,-0.5)
\psarc[linewidth=0.5pt]{-}(-0.5,-0.5){0.25}{0}{90}
\rput(0,0){$u$} \unlw
\end{pspicture}\ =\ \cos u\ \ 
\begin{pspicture}(-0.5,-0.1)(0.5,0.5) \lw
\psline{-}(-0.5,-0.5)(0.5,-0.5)(0.5,0.5)(-0.5,0.5)(-0.5,-0.5)\unlw
\psset{linecolor=myc}
\psarc{-}(0.5,-0.5){0.5}{90}{180}
\psarc{-}(-0.5,0.5){0.5}{270}{360}
\end{pspicture}\ +\ \sin u\ \ 
\begin{pspicture}(-0.5,-0.1)(0.5,0.5) \lw
\psline{-}(-0.5,-0.5)(0.5,-0.5)(0.5,0.5)(-0.5,0.5)(-0.5,-0.5)\unlw
\psset{linecolor=myc}
\psarc{-}(-0.5,-0.5){0.5}{0}{90}
\psarc{-}(0.5,0.5){0.5}{180}{270}
\end{pspicture}
\ \ =\ \ 
\begin{pspicture}(-0.5,-0.1)(0.5,0.5)\lw
\psline{-}(-0.5,-0.5)(0.5,-0.5)(0.5,0.5)(-0.5,0.5)(-0.5,-0.5) \unlw
\psarc[linewidth=0.5pt]{-}(0.5,-0.5){0.25}{90}{180}
\rput(0,0){$\frac\pi2-u$}
\end{pspicture}
\end{equation*}
\\
and $u\in[0,\frac\pi2]$ is the anisotropy parameter. (A definition of $D_N(u)$ for general $\beta$ exists, see \cite{PRZ}.) From the definition, it can easily be shown that $D_N(u) = D_N(\pi/2-u)$ and $D_N (0) = D_N(\pi/2) = id$ are satisfied, where $id$ is the unique connectivity connecting every point on top to the corresponding point on the bottom.  In \cite{PR}, it is also shown that $D_N(u)$ satisfies the following inversion identity:
\begin{equation*}D_N(u)D_N(u+\frac{\pi}{2}) = \left( \frac{\cos^{2N} u - \sin^{2N} u}{\cos^{2} u - \sin^{2} u}\right)^2 id,\end{equation*}
from which is it possible to retrieve a closed expression for the eigenvalues of $D_N(u)$, which we note $d_N(u)$:
\begin{align} 
\textrm{N odd:} \qquad d_N(u) &= \frac{1}{2^{N-1}} \prod_{j=1}^{\frac{N-1}{2}} \left( \frac1{\sin \frac{(2j-1)\pi}{2N}} + \epsilon_j \sin 2u \right) \left( \frac1{\sin \frac{(2j-1)\pi}{2N}} + \mu_j \sin 2u \right),   \label{eq:dNimpair} \\
\textrm{N even:} \qquad d_N(u) &= \frac{N}{2^{N-1}} \prod_{j=1}^{\frac{N-2}{2}} \left( \frac1{\sin \frac{j\pi}{N}} + \epsilon_j \sin 2u \right) \left( \frac1{\sin \frac{j\pi}{N}} + \mu_j \sin 2u \right),  \label{eq:dNpair}
\end{align}
where $\epsilon_j,\mu_j = \pm 1$ for every $j$. Fixing values for each $\epsilon_j$ and each $\mu_j$, the set of zeroes of $d_N(u)$ is
$$\{ u|{d_N(u)=0} \}= \bigcup_{\nu = \epsilon, \mu }\bigcup_j \, \Big\{ (2+\nu_j)\frac{\pi}4 \pm \frac{i}2 \ln \tan \frac{t_j}2 + \pi k ,  \quad k \in \mathbb{Z} \Big\}$$
where
\begin{align*}
\textrm{N odd:} \qquad t_j &= \frac{(2j-1)\pi}{2N}, \\
\textrm{N even:} \qquad t_j &= \frac{j\pi}N .
\end{align*}

Given a fixed $d_N(u)$, every zero in the above set appears $0$, $1$ or $2$ times, and the number of zeroes with imaginary value $ i/2 \ln \tan t_j/2$ is always $2$. There are ${N-1}$ zeroes for $N$ odd and ${N-2}$ for $N$ even, which results in a total of $2^{N-1}$ and $2^{N-2}$ choices, respectively, for the eigenvalues $d_N(u)$. The set of possible solutions for eigenvalues of $\rho(D_N(u))$ is too large and one must identity which ones are {\it relevant}. This will be the subject of the next section.
\\
\\
 $D_N(u)$ can be developed in a Taylor series around the point $u=0$, yielding
\begin{equation}D_N(u) =id + 2u \mathcal{H}_N +o(u^2)\qquad \textrm{with} \qquad \mathcal{H}_N = \sum_{i=1}^{N-1} U_i.\label{eq:DetH}\end{equation}
To understand and prove the selection rules, we will calculate the eigenvalues of $\mathcal{H}_N$. Using the expansions of (\ref{eq:dNimpair}) and (\ref{eq:dNpair}) around $u=0$, and using $d_N(0)=1$, i.e.
$$  \frac{1}{2^{N-1}} \prod_{j=1}^{\frac{N-1}2} \frac{1}{\sin^2 \frac{(2j-1)\pi}{2N}} = 1 \qquad \textrm{and} \qquad \frac{N}{2^{N-1}} \prod_{j=1}^{\frac{N-2}2} \frac{1}{\sin^2 \frac{j\pi}{N}} = 1$$
for $N$ odd and $N$ even respectively, one finds that eigenvalues of $\mathcal{H}_N$, denoted $h_N$, are
\begin{align}
\textrm{N odd:} \qquad h_N &= \sum_{j=1}^{\frac{N-1}2} \cos \left( \frac{\pi j}N \right) (\epsilon_{\frac{N+1}2 - j} + \mu_{\frac{N+1}2 - j}),  \label{eq:hNimpair}\\
\textrm{N even:} \qquad h_N &= \sum_{j=1}^{\frac{N-2}2} \cos \left( \frac{\pi j}N \right) (\epsilon_{\frac{N}2 - j} + \mu_{\frac{N}2 - j}),\label{eq:hNpair}
\end{align}
and the $\epsilon_j$s and $\mu_j$s are those of $d_N(u)$. 

\subsection{Two-column configurations}\label{sec:2column}

The selection rules given in \cite{PR} have been formulated in terms of column configurations. This section is a quick review of their definitions. 
\begin{Definition} A one-column configuration of height $M$ is a configuration of $M$ sites disposed in a column and labeled from $1$ to $M$, starting from the top. In a column configuration, every site is either occupied or unoccupied and we define its signature, $S = \{S_1, S_2, ..., S_m\}$, where the $S_i$s are the labels of the occupied sites in ascending order (and $m\le M$ is their number and will be called the length of the signature). We identify unoccupied sites with white circles ``\, \,
\psset{unit=0.6}
\begin{picture}(-2,-3.5)(6.0,-3.0)
\pscircle[fillstyle=solid,fillcolor=white,linewidth=1pt](0,0){0.175}
\psset{linewidth=1pt}
\end{picture}''
and occupied sites with blue 
 circles ``\, \,
\begin{picture}(-2,-3.5)(6.0,-3.0)
\pscircle[fillstyle=solid,fillcolor=blue,linewidth=1pt](0,0){0.175}
\psset{linewidth=1pt}
\end{picture}''.
\end{Definition}

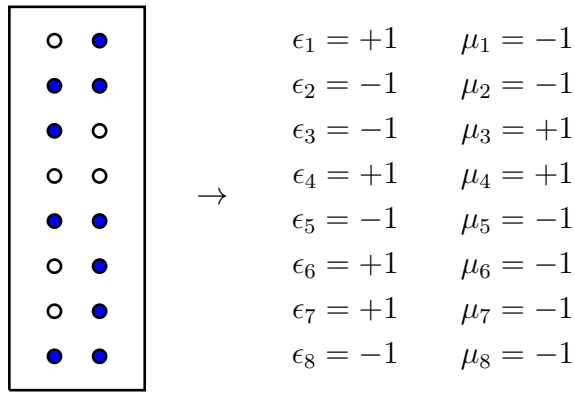
\begin{figure}[h!]
\begin{center}
\psset{unit=0.6}
\begin{pspicture}(0,-0.5)(14.5,9)
\psset{linewidth=1pt}
\rput(4.5,4.5){$\rightarrow$}
\rput(9.5,1){$\epsilon_8 = -1 \qquad \mu_8 = -1$ }
\rput(9.5,2){$\epsilon_7 = +1 \qquad \mu_7 = -1$ }
\rput(9.5,3){$\epsilon_6 = +1 \qquad \mu_6 = -1$ }
\rput(9.5,4){$\epsilon_5 = -1 \qquad \mu_5 = -1$ }
\rput(9.5,5){$\epsilon_4 = +1 \qquad \mu_4 = +1$ }
\rput(9.5,6){$\epsilon_3 = -1 \qquad \mu_3 = +1$ }
\rput(9.5,7){$\epsilon_2 = -1 \qquad \mu_2 = -1$ }
\rput(9.5,8){$\epsilon_1 = +1 \qquad \mu_1 = -1$ }
\pscircle[fillstyle=solid,fillcolor=blue,linewidth=1pt](1,1){0.175}\pscircle[fillstyle=solid,fillcolor=blue,linewidth=1pt](2,1){0.175}
\pscircle[fillstyle=solid,fillcolor=white,linewidth=1pt](1,2){0.175}\pscircle[fillstyle=solid,fillcolor=blue,linewidth=1pt](2,2){0.175}
\pscircle[fillstyle=solid,fillcolor=white,linewidth=1pt](1,3){0.175}\pscircle[fillstyle=solid,fillcolor=blue,linewidth=1pt](2,3){0.175}
\pscircle[fillstyle=solid,fillcolor=blue,linewidth=1pt](1,4){0.175}\pscircle[fillstyle=solid,fillcolor=blue,linewidth=1pt](2,4){0.175}
\pscircle[fillstyle=solid,fillcolor=white,linewidth=1pt](1,5){0.175}\pscircle[fillstyle=solid,fillcolor=white,linewidth=1pt](2,5){0.175}
\pscircle[fillstyle=solid,fillcolor=blue,linewidth=1pt](1,6){0.175}\pscircle[fillstyle=solid,fillcolor=white,linewidth=1pt](2,6){0.175}
\pscircle[fillstyle=solid,fillcolor=blue,linewidth=1pt](1,7){0.175}\pscircle[fillstyle=solid,fillcolor=blue,linewidth=1pt](2,7){0.175}
\pscircle[fillstyle=solid,fillcolor=white,linewidth=1pt](1,8){0.175}\pscircle[fillstyle=solid,fillcolor=blue,linewidth=1pt](2,8){0.175}
\psline{-}(0,0.25)(3,0.25)(3,8.75)(0,8.75)(0,0.25)
\end{pspicture}
\caption{An admissible two-column configuration in $A^8_{4,6}$ with $L=(2,3,5,8)$ and $R=(1,2,5,6,7,8)$: blue 
sites are occupied and white sites unoccupied. To its right, the corresponding values of the $\epsilon_j$s and $\mu_j$s.} \label{fig:bij}
\end{center}
\end{figure}

\begin{Definition} A two-column configuration of height $M$ is a pair of one-column configurations, both of height $M$, and is usually depicted as in Figure \ref{fig:bij}. Its signature is $S = (L,R)$, where $L$ and $R$ are the respective signatures of the left and right column configurations and may have different lengths $m$ and $n$. A two-column configuration will be said to be {\it admissible} if $0 \le m \le n \le M$ and $L_i \ge R_i$ for all $i = 1, ..., m$.  We denote by $A^M_{m,n}$ the set of admissible two-column configurations of height $M$ and signature lengths $m$ and $n$. When $m$, $n$ and $M$ are such that the previous constraint is violated, $A^M_{m,n} \equiv \emptyset$.
\end{Definition}

The graphical interpretation of this last definition is simple. Fix a two-column configuration. To see if it is admissible, we draw on the two-column configuration segments connecting sites with label $L_i$ from the left column to sites with label $R_i$ from the left column, for $i = 1, ..., m$ (the remaining sites at positions $R_j$ with $m<j\le n$ are not connected to any other site). If all the segments have positive or null slopes, the configuration is admissible.

\begin{Definition}
The {\it reduced} set $\tilde{A}^{x+y}_{x,y}$ of admissible two-column configurations is the subset of  configurations  of $A^{x+y}_{x,y}$ that have one and only one excitation for every $j$.
\label{sec:deuxcolreduit}
\end{Definition}
Evaluating $|\tilde{A}^{x+y}_{x,y}|$ is simple, as there exists bijections between reduced configurations in $\tilde{A}^{x+y}_{x,y}$, Dyck paths  $\vec x \in DP^{x+y}_{y-x}$ (see definition \ref{sec:Dyckpath}) and link states in $V_{x+y}^{y-x}$:
\begin{itemize}
\item From an element of $\tilde{A}^{x+y}_{x,y}$, we set $\epsilon_j = +1$ if the site of the left one-column configuration at height $j$ is unoccupied, and $-1$ otherwise. $\vec x = (\epsilon_1, ..., \epsilon_{x+y})$ is a Dyck path of length $x+y$ as, from the definition of reduced admissible configurations, $\sum_{i = 1}^k \epsilon_i \ge 0$ for every $k$ in $1, ..., x+y$. Since there are, in total, $y$ ``$+1$''s and $x$ ``$-1$''s, the endpoint of the Dyck Path is at $y-x$. This transformation is bijective.
\item The bijection between Dyck paths and link states is given by the following. To each of the entries of the link state, we associate the integer $j$ in $1, ..., N$ from left to right and build pairings $(j',j)$ (the positions where the bubbles connect). Starting from the left, for every $x_j = -1$, we pair $j$ to the closest available $j'$ such that $x_{j'}=+1$ and $j>j'$. When every $j$ with $x_j = -1$ is paired, the remaining $y-x$ unpaired sites are chosen to be defects. The link state $v$ obtained from a given Dyck path $\vec x$ by the previous procedure will be noted $v = \mathcal{B}(\vec{x})$. 

\end{itemize}
From this bijection, 
\begin{equation}|\tilde{A}^{x+y}_{x,y}| = \textrm{dim}V_{x+y}^{y-x} =  \begin{pmatrix} x+y \\ x \end{pmatrix} - \begin{pmatrix} x+y \\ x-1 \end{pmatrix}.  \label{eq:sizeAtilde}\end{equation}
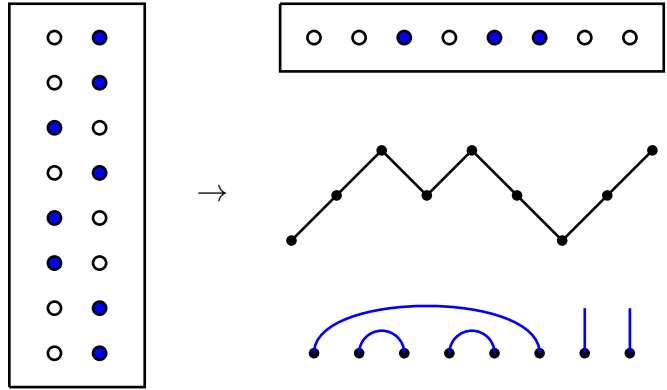
\begin{figure}[h!]
\begin{center}
\psset{unit=0.6}
\begin{pspicture}(0,-0.5)(14.5,9)
\psset{linewidth=1pt}
\pscircle[fillstyle=solid,fillcolor=white,linewidth=1pt](1,1){0.175}\pscircle[fillstyle=solid,fillcolor=blue,linewidth=1pt](2,1){0.175}
\pscircle[fillstyle=solid,fillcolor=white,linewidth=1pt](1,2){0.175}\pscircle[fillstyle=solid,fillcolor=blue,linewidth=1pt](2,2){0.175}
\pscircle[fillstyle=solid,fillcolor=blue,linewidth=1pt](1,3){0.175}\pscircle[fillstyle=solid,fillcolor=white,linewidth=1pt](2,3){0.175}
\pscircle[fillstyle=solid,fillcolor=blue,linewidth=1pt](1,4){0.175}\pscircle[fillstyle=solid,fillcolor=white,linewidth=1pt](2,4){0.175}
\pscircle[fillstyle=solid,fillcolor=white,linewidth=1pt](1,5){0.175}\pscircle[fillstyle=solid,fillcolor=blue,linewidth=1pt](2,5){0.175}
\pscircle[fillstyle=solid,fillcolor=blue,linewidth=1pt](1,6){0.175}\pscircle[fillstyle=solid,fillcolor=white,linewidth=1pt](2,6){0.175}
\pscircle[fillstyle=solid,fillcolor=white,linewidth=1pt](1,7){0.175}\pscircle[fillstyle=solid,fillcolor=blue,linewidth=1pt](2,7){0.175}
\pscircle[fillstyle=solid,fillcolor=white,linewidth=1pt](1,8){0.175}\pscircle[fillstyle=solid,fillcolor=blue,linewidth=1pt](2,8){0.175}
\psline{-}(0,0.25)(3,0.25)(3,8.75)(0,8.75)(0,0.25)
\rput(4.5,4.5){$\rightarrow$}
\psline{-}(6,8.75)(14.50,8.75)(14.50,7.25)(6,7.25)(6,8.75)
\pscircle[fillstyle=solid,fillcolor=white,linewidth=1pt](6.75,8){0.175}
\pscircle[fillstyle=solid,fillcolor=white,linewidth=1pt](7.75,8){0.175}
\pscircle[fillstyle=solid,fillcolor=blue,linewidth=1pt](8.75,8){0.175}
\pscircle[fillstyle=solid,fillcolor=white,linewidth=1pt](9.75,8){0.175}
\pscircle[fillstyle=solid,fillcolor=blue,linewidth=1pt](10.75,8){0.175}
\pscircle[fillstyle=solid,fillcolor=blue,linewidth=1pt](11.75,8){0.175}
\pscircle[fillstyle=solid,fillcolor=white,linewidth=1pt](12.75,8){0.175}
\pscircle[fillstyle=solid,fillcolor=white,linewidth=1pt](13.75,8){0.175}
\psline{-}(6.25,3.5)(7.25,4.5)(8.25,5.5)(9.25,4.5)(10.25,5.5)(11.25,4.5)(12.25,3.5)(13.25,4.5)(14.25,5.5)
\psdots(6.25,3.5)(7.25,4.5)(8.25,5.5)(9.25,4.5)(10.25,5.5)(11.25,4.5)(12.25,3.5)(13.25,4.5)(14.25,5.5)
\psdots(6.75,1)(7.75,1)(8.75,1)(9.75,1)(10.75,1)(11.75,1)(12.75,1)(13.75,1)
\psset{linecolor=myc2}
\psarc{-}(8.25,1){0.5}{0}{180}
\psarc{-}(10.25,1){0.5}{0}{180}
\psbezier{-}(6.75,1)(6.75,2.4)(11.75,2.4)(11.75,1)
\psline{-}(12.75,1)(12.75,2)
\psline{-}(13.75,1)(13.75,2)
\end{pspicture}
\caption{ A two-column admissible configuration in $\tilde A^8_{3,5}$ and, to the right, the corresponding Dyck path $\in DP^8_2$ and  link state $\in V^2_8$.} \label{fig:columnconfig}
\end{center}
\end{figure}

\subsection{Conjectured degeneracies and selection rules}\label{sec:selection}
In this section, we state the conjecture of \cite{PR} and use the definitions of $A^M_{m,n}$ to translate it in terms of degeneracies in the spectrum of $\rho(\mathcal H_N(u))$. To each two-column configuration corresponds a choice of  $\epsilon_j$ and $\mu_j$. The rules are the following :
\begin{itemize}
\item A white circle `` \,
\psset{unit=0.6}
\begin{picture}(-2,-3.5)(6.0,-3.0)
\pscircle[fillstyle=solid,fillcolor=white,linewidth=1pt](0,0){0.175}
\psset{linewidth=1pt}
\end{picture}''
corresponds to $+1$ and a blue 
 circle `` \,
\begin{picture}(-2,-3.5)(6.0,-3.0)
\pscircle[fillstyle=solid,fillcolor=blue,linewidth=1pt](0,0){0.175}
\psset{linewidth=1pt}
\end{picture}''
to a $-1$.
\item The left column corresponds to $\epsilon$ excitations, and the right to $\mu$ excitations.
\item As before, $j$ grows from top to bottom.
\end{itemize}
Pearce and Rasmussen \cite{PR} give the following conjecture:
\begin{Conjecture}
 In the sector with $d$ defects, the set of choices of the $\epsilon_j$s and $\mu_j$s belonging to
\begin{equation}
N \, \textrm{odd:} \quad \bigcup_{p=0}^{\frac{N-d}2} A^{\frac{N-1}2}_{p,p+\frac{d-1}2}, \qquad \qquad N \, \textrm{even:} \quad \bigcup_{p=0}^{\frac{N-d}2} \left( A^{\frac{N-2}2}_{p,p+\frac{d-2}2} \cup A^{\frac{N-2}2}_{p,p+\frac{d}2} \right), \label{eq:conjecture}
\end{equation}
forms the spectrums of $\rho(D_N(u))$ and $\rho(\mathcal H_N)$. \label{sec:conjec}\end{Conjecture}
Recall that when some indices of $A^M_{m,n}$ do not satisfy the constraint $0 \le m \le n \le M$, the set $A^M_{m,n}$ is empty. In this sense, the case $d=0$ is special, as the selection rule reduces to 
\begin{equation} \bigcup_{p=0}^{\frac{N-2}2} A_{p,p}^{\frac{N-2}2}.  \label{eq:d=0}\end{equation}

\begin{Definition}
The set of eigenvalues of $\rho(\mathcal{H}_N)$ in the sector with $d$ defects, as given by the selection rules (\ref{eq:conjecture}), will be noted $ H^d_N$. An eigenvalue $\lambda$ will be said to belong to $ A^{M}_{m,n}$ if it can be obtained by a choice of $\epsilon_j$s and $\mu_j$s represented by an element of $A^{M}_{m,n}$. For $N$ even, we distinguish between $H_{N,0}^d$ and  $H_{N,1}^d$, the sets of eigenvalues $\lambda$ obtained from admissible two-column configurations in $ \cup_{p=0}^{\frac{N-d}2}  A^{\frac{N-2}2}_{p,p+\frac{d-2}2} $ and $ \cup_{p=0}^{\frac{N-d}2}  A^{\frac{N-2}2}_{p,p+\frac{d}2} $ respectively.
\end{Definition}

In the following, the cases $N$ odd and $N$ even will often be treated separately. In preparation, we give the following two definitions.

\begin{Definition}
Let $\delta=0,1$,  we define the set  $ \Lambda_\delta^n$ of $\lambda$s given by
\begin{equation} \lambda= 2\sum_{i=1}^{m} \eta_{k_i} \cos \frac{\pi k_i}N, \qquad \textrm{where}\label{eq:lambdaint}\end{equation}
\begin{itemize}
\item $\eta_i = \pm 1$ for all $i$;
\item $m$ may take all values satisfying both $0 \le m \le n$ and $n-m \equiv \delta \mod 2$;
\item $k_i \in \mathbb{N} $, \quad $1 \le k_1 < k_2 < ... < k_m \le F(N) \quad \textrm{with} \quad F(N) =\left\{ \begin{array}{l l} (N-1)/2, & \quad N \textrm{odd},\\ (N-2)/2,& \quad N \textrm{even}. \end{array} \right.$
\end{itemize}
Let $\lambda \in  \Lambda_\delta^n $. We also define
\begin{itemize}
\item $ K^+ $ : the set of $k$s in $\{ k_1,...,k_m \}$ with $\eta_{k_i} = +1$,
\item $ K^- $ : the set of $k$s in  $\{k_1,...,k_m\} $ with $\eta_{k_i} = -1$,
\item $ K^c $ : the set of $k$s in $\{1, ..., F(N)\}$ that are neither in $K^+$ nor $K^-$.
\end{itemize}
\label{sec:enslambimpair}
\end{Definition}

To each $\lambda \in \Lambda^n_\delta $ we associate the smallest number  $m$ such that $\lambda$ can be written as (\ref{eq:lambdaint}), ignoring accidental cancellations. For instance, with $N=9$, $\lambda_1 = 0$ has $m=0$ and $\lambda_2 =  2 \cos{\pi/9} -2 \cos{2 \pi/9} - 2 \cos{4\pi/9}$ has  $m=3$, even though $\lambda_2$ evaluates to $0$. The accidental degeneracies like the one given previously will not be considered, as they are  degeneracies of $\rho(\mathcal{H}_N)$, but not of $\rho(D_N(u))$.

\subsection{$N$ odd}

\begin{Proposition}
The two sets $H_N^d$ and $\Lambda_0^{(N-d)/2}$ are equal.
\label{sec:TLinint}
\end{Proposition}
\noindent{\scshape Proof\ \ } 
First, let $h \in H_N^d $. It is obvious that $h$ can be written as (\ref{eq:lambdaint}), for a certain $0 \le m \le (N-1)/2$. Here are the rules: if at level $j$
\begin{itemize}
\item[(a)]  there are two white circles,  put $k_j$ in $K^+$;
\item[(b)]  there are two blue 
circles, put $k_j$ in $K^-$;
\item[(c)]  there is one white and one blue circle, put $k_j$ in $K^c$.
\end{itemize}

To prove that $h \in \Lambda_0^{(N-d)/2}$, one must show two things: that the top bound for $m$ can be lowered from $(N-1)/2$ to $(N-d)/2$, and that $n-m = 0 \, \textrm{mod} \, 2$.  To do this, note first that if $h \in A^{(N-1)/2}_{p,p+(d-1)/2}$, the maximal number of elements in $K^-$ and $K^+$ are $p$ and $(N-d)/2-p$ respectively (and these two events occur simultaneously). The maximal value of $m \equiv |K^+ \cup K^-|$ is $(N-d)/2$; it never goes beyond $n$. The values $m$ can take make jumps of $2$ and are  $n, n-2, n-4, ..., 0$:  $n-m = 0 \, \textrm{mod} \, 2$ as expected.
\\

\noindent Second, let $\lambda \in  \Lambda^n_0$ with $m$ fixed. We show that $\lambda \in  H_N^{N-2n}$. The rule is the following:
\begin{itemize}
\item[(a)] if $k_j \in K^+$, put two white circles at level $j$; 
\item[(b)] if $k_j \in K^- $, put two blue circles at level $j$; 
\item[(c)] if $k_j \in K^c $, put one circle of each color at level $j$.
\end{itemize} 
One must then choose carefully the position of the pairs of colored circles in $(c)$, to ensure that the two-column configuration is admissible and that it is in $A^{(N-1)/2}_{p,p+(d-1)/2}$ for some $p$. Among all $k_j$ in $K^c $, one must put $a_1$ blue circles in the left column and $a_2$ in the right column, and impose that $a_1+a_2 = |K^c| =  (N-1)/2-m$ and $a_2-a_1 = (N-1)/2-n$. This is always possible, with the choice $a_1=(n-m)/2$ and  $a_2 =(N-n-m-1)/2$ (note that $a_1$ and $a_2$ are integers). $\lambda$ is then contained in $A_{p,p+(d-1)/2}^{(N-1)/2}$ with $p=|K^-|+(n-m)/2$.
\hfill$\square$\\ 

From the previous proof, all the eigenvalues of $\rho(\mathcal{H}_N)$ are in $\Lambda^n_0$, and we need not worry about values in $\Lambda^n_1$. For a given element of $\Lambda^n_0$, we can now calculate its degeneracy in $ \rho(\mathcal{H}_N)$ in the sector with $d$ defects, as given by the selection rules. The following statement is therefore equivalent to conjecture \ref{sec:conjec} for N odd (omitting accidental degeneracies):

\begin{Conjecture}
Let $\lambda \in \Lambda^n_0 $ with a fixed value of $m$ (and $n-m \equiv 0 \mod 2$). Its degeneracy in $\rho(\mathcal{H}_N)$ in the sector with $N-2n$ defects, as conjectured in \cite{PR}, is
\begin{equation}\textrm{deg}_{\mathcal{H}}(\lambda) = \begin{pmatrix} \frac{N-1}2 -m \\ \frac{n-m}{2} \end{pmatrix} - \begin{pmatrix} \frac{N-1}2 -m \\ \frac{n-m-2}2 \end{pmatrix}, \qquad 0 \le m \le n.
\label{eq:conjequiv}\end{equation}
\label{sec:degimpair}\end{Conjecture}
\noindent{\scshape Proof\ \ } 
In the second part of the previous proof, for every $k_j$ in $K^c $, there was a freedom in the choice of admissible configurations. To count the degeneracies, one has to count these possible choices, as a pair of occupied and unoccupied sites at height $j$ gives contribution $0$ to eigenvalues of $\rho(\mathcal H_N)$, regardless of $j$. For a given two-column configuration, whether it is admissible does not depend on levels with two blue circles or two white circles. These can be removed.
The configuration resulting from this operation is in the reduced set $\tilde{A}^{(N-1)/2-m}_{(n-m)/2,(N-1-n-m)/2}$ whose dimension, given by (\ref{eq:sizeAtilde}), is the desired result (\ref{eq:conjequiv}). \hfill$\square$  \\

\begin{figure}[h!]
\begin{center}
\psset{unit=0.6}
\begin{pspicture}(0,-0.5)(5.75,9)
\psset{linewidth=1pt}
\rput(4.5,4.5){$\rightarrow$}
\pscircle[fillstyle=solid,fillcolor=blue,linewidth=1pt](1,1){0.175}\pscircle[fillstyle=solid,fillcolor=blue,linewidth=1pt](2,1){0.175}
\pscircle[fillstyle=solid,fillcolor=white,linewidth=1pt](1,2){0.175}\pscircle[fillstyle=solid,fillcolor=blue,linewidth=1pt](2,2){0.175}
\pscircle[fillstyle=solid,fillcolor=white,linewidth=1pt](1,3){0.175}\pscircle[fillstyle=solid,fillcolor=blue,linewidth=1pt](2,3){0.175}
\pscircle[fillstyle=solid,fillcolor=blue,linewidth=1pt](1,4){0.175}\pscircle[fillstyle=solid,fillcolor=blue,linewidth=1pt](2,4){0.175}
\pscircle[fillstyle=solid,fillcolor=white,linewidth=1pt](1,5){0.175}\pscircle[fillstyle=solid,fillcolor=white,linewidth=1pt](2,5){0.175}
\pscircle[fillstyle=solid,fillcolor=blue,linewidth=1pt](1,6){0.175}\pscircle[fillstyle=solid,fillcolor=white,linewidth=1pt](2,6){0.175}
\pscircle[fillstyle=solid,fillcolor=blue,linewidth=1pt](1,7){0.175}\pscircle[fillstyle=solid,fillcolor=blue,linewidth=1pt](2,7){0.175}
\pscircle[fillstyle=solid,fillcolor=white,linewidth=1pt](1,8){0.175}\pscircle[fillstyle=solid,fillcolor=blue,linewidth=1pt](2,8){0.175}
\psline{-}(0,0.25)(3,0.25)(3,8.75)(0,8.75)(0,0.25)
\end{pspicture}
\begin{pspicture}(0,-0.5)(4.5,9)
\psset{linewidth=1pt}
\pscircle[fillstyle=solid,fillcolor=white,linewidth=1pt](1,2){0.175}\pscircle[fillstyle=solid,fillcolor=blue,linewidth=1pt](2,2){0.175}
\pscircle[fillstyle=solid,fillcolor=white,linewidth=1pt](1,3){0.175}\pscircle[fillstyle=solid,fillcolor=blue,linewidth=1pt](2,3){0.175}
\pscircle[fillstyle=solid,fillcolor=blue,linewidth=1pt](1,6){0.175}\pscircle[fillstyle=solid,fillcolor=white,linewidth=1pt](2,6){0.175}
\pscircle[fillstyle=solid,fillcolor=white,linewidth=1pt](1,8){0.175}\pscircle[fillstyle=solid,fillcolor=blue,linewidth=1pt](2,8){0.175}
\psline{-}(0,0.25)(3,0.25)(3,8.75)(0,8.75)(0,0.25)
\end{pspicture}
\caption{A two-column admissible configuration in $ A^8_{4,6}$ and its corresponding reduced configuration in $\tilde{A}^4_{1,3}$. It corresponds to the eigenvalue $-2 \cos \pi/17 -2 \cos 4\pi/17 +2 \cos 5\pi/17 -2 \cos 7\pi/17$ of $\rho(\mathcal H_{N=17})$ and has degeneracy $3$.} \label{fig:redu}
\end{center}
\end{figure}
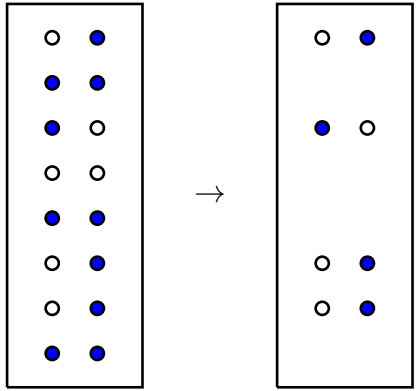

\subsection{$N$ even}

The case $N$ even is analogous to the case $N$ odd, though the selection rule is more complicated.

\begin{Proposition} Let $\delta = 0,1$. Then $H_{N,\delta}^d = \Lambda_\delta^{(N-d)/2}$.
\end{Proposition}
\noindent{\scshape Proof\ \ } We start by showing that for $\delta = 0, 1$, $H_{N,\delta}^d \subset \Lambda_\delta^{(N-d)/2}$. The beginning of this proof is identical to that of proposition \ref{sec:TLinint}. The arguments for lowering the upper bound for $m$ from $(N-2)/2$ to $(N-d)/2$ and for the parity of $n-m$ must be repeated. (Note that in the case $d=0$, it seems that this raises the upper bound, but since the selection rule is given in (\ref{eq:d=0}), this is not the case.) For $\delta = 0$, $K^-$ has at most $p$ elements and $ K^+ $, at most $(N-d)/2-p$. Then, $m = |K^+  \cup  K^-|$ is at most $n = (N-d)/2$ and $m$ takes values $n, n-2, ...$; this is the case $n-m = 0 \,\textrm{mod} \, 2$. For $\delta = 1$, $\textrm{max} \, \, m = \textrm{max} | K^- | = p$, $\textrm{max} |  K^+ | = (N-d)/2 -p - 1$, $\textrm{max} (| K^+  \cup   K^- |) = (N-d)/2 - 1 = n-1$, and $n-m = 1\, \textrm{mod} \,2$.

In the other direction, we show $\Lambda_0^{(N-d)/2} \subset H_{N,0}^d$. The rules are those of proposition \ref{sec:TLinint}. The positions of the pairs in $K^c$ is as follows:
\begin{itemize}
\item If $\lambda \in \Lambda^n_0 $, the constraints are $a_1+a_2 = (N-2)/2-m$ and $a_2-a_1 = (N-2)/2-n$. Among the $k_j$s in $K^c$, we put $a_1=(n-m)/2$ excitations in the left column and $a_2=(N-n-m-2)/2$ in the right column.
\item If $\lambda \in  \Lambda^n_1 $, the constraints are $a_1+a_2 = (N-2)/2-m$ and $a_2-a_1 = N/2-n$. Among the $k_j$s in $K^c$, we put $a_1=(n-m-1)/2$ excitations in the left column and $a_2=(N-n-m-1)/2$ in the right column.
\end{itemize}
\hfill$\square$  

For $N$ even, the following is the translation of the conjecture \ref{sec:conjec}:
\begin{Conjecture}
The conjectured degeneracy of $\lambda \in \Lambda^n_\delta $, $m$ fixed (and $n-m \equiv \delta \mod 2$), in the sector $d= N-2n$, is given by
\begin{align} 
\delta=0: \qquad \textrm{deg}_{\mathcal{H}}(\lambda) &= \begin{pmatrix} \frac{N-2}2 -m \\ \frac{n-m}{2} \end{pmatrix} - \begin{pmatrix} \frac{N-2}2 -m \\ \frac{n-m-2}2 \end{pmatrix}, \qquad 0 \le m \le n, \label{eq:conjequiv2} \\
\delta=1: \qquad \textrm{deg}_{\mathcal{H}}(\lambda) &= \begin{pmatrix} \frac{N-2}2 -m \\ \frac{n-m-1}{2} \end{pmatrix} - \begin{pmatrix} \frac{N-2}2 -m \\ \frac{n-m-3}2\end{pmatrix}, \qquad 0 \le m \le n-1.
\label{eq:conjequiv3}\end{align}
\label{sec:degpair}
\end{Conjecture}
This proof is identical to that of \ref{sec:degimpair} and left to the reader. One can also verify that these formulae are valid for $d=0$ and that $\textrm{deg}_{\mathcal{H}}(\lambda) = 0$ for $\delta = 0$, as expected. The results of the conjectures \ref{sec:degimpair} and \ref{sec:degpair} are statements equivalent to (\ref{eq:conjecture}): they provide a conjecture for degeneracies of eigenvalues of $\rho(\mathcal{H}_N)$ in the sector with $d = N-2n$ defects (in fact, the statement is not as strong because of the accidental degeneracies due to exceptional trigonometric identities, but these will be ignored). To prove the selection rules, we will show that degeneracies of $\rho(\mathcal{H}_N)$ are indeed given by eqs (\ref{eq:conjequiv}),  (\ref{eq:conjequiv2}) and (\ref{eq:conjequiv3}).

%
\section{The XXZ Hamiltonian} \label{sec:XXZ}
%

\noindent On the finite (non-periodic) lattice, the well-studied \cite{PasquierSaleur} XXZ Hamiltonian for spin-$\frac12$ particles is
\begin{equation}
H^q_{XXZ} = \frac12 \left( \sum_{j=1}^{N-1} ( \sigma^x_j\sigma^x_{j+1} + \sigma^y_j\sigma^y_{j+1} + \frac{q+q^{-1}}2 \sigma^z_j\sigma^z_{j+1} ) -  \frac{q-q^{-1}}2 \left( \sigma^z_1 - \sigma^z_{N} \right) \right),
\label{eq:HXXZ}
\end{equation}
where 
$$ \sigma^a_j = \underbrace{id_2 \otimes id_2 \otimes \dots \otimes id_2}_{j-1} \otimes \, \sigma^a \otimes \underbrace{id_2 \otimes id_2 \otimes \dots \otimes id_2}_{N-j}.$$
This Hamiltonian acts on $(\mathbb{C}^2)^{\otimes N}$ and can also be written as
$$H_{XXZ}^q = \sum_{j=1}^{N-1} \left( \frac{q+q^{-1}}4 I + e_i \right), \qquad \textrm{where}$$
\begin{align} e_j &= \frac12 \left( \sigma_j^x \sigma_{j+1}^x + \sigma_j^y \sigma_{j+1}^y + \frac{q+q^{-1}}2 (\sigma_j^z \sigma_{j+1}^z - id) - \frac{q-q^{-1}}2 (\sigma_j^z - \sigma_{j+1}^z) \right) \nonumber \\
&= \underbrace{id_2 \otimes id_2 \otimes \dots \otimes id_2}_{j-1} \otimes \, \tilde{e} \otimes \underbrace{id_2 \otimes id_2 \otimes \dots \otimes id_2}_{N-j-1} 
\label{eq:eTL}\end{align}
\begin{equation}\textrm{and} \qquad  \tilde{e} = \begin{pmatrix} 
0 & 0 & 0 & 0 \\
0 & -q & 1 & 0 \\
0 & 1 & -q^{-1} & 0 \\
0 & 0 & 0 & 0
\end{pmatrix}.\label{eq:etilde}\end{equation}
The matrices $e_j$s form a representation of $TL_N(\beta)$ with $\beta = -(q+q^{-1})$. We will be interested in diagonalizing this Hamiltonian when $q=i$. More precisely, we will show that $H^{q=i}_{XXZ}$ can be diagonalized when $N$ is odd, but not when $N$ is even, in which case we shall give its Jordan form. We start with

\begin{equation*}
H \equiv H^{q=i}_{XXZ} = \frac12 \left( \sum_{j=1}^{N-1}  (\sigma^x_j\sigma^x_{j+1} + \sigma^y_j\sigma^y_{j+1}  ) - i \left( \sigma^z_1 - \sigma^z_{N} \right) \right).
\end{equation*}

\subsection{Free fermions}
\noindent Ideas in this section are similar to those found in \cite{LSM}, \cite{Deguchi} and \cite{Bilstein}. $H$ can be transformed by writing $\sigma_j^x$, $\sigma_j^y$ and $\sigma_j^z$ in terms of  $ \sigma^\pm_j = (\sigma^x_j \pm i \sigma^y_j)/2 $:

\begin{equation*}
H =  \sum_{j=1}^{N-1} \left( \sigma^+_j\sigma^-_{j+1} + \sigma^-_j\sigma^+_{j+1}  \right) -i (\sigma^+_1\sigma^-_{1}-\sigma^+_N\sigma^-_{N}). 
\end{equation*}

\noindent We perform the celebrated Jordan-Wigner transformation by passing to creation and annihilation operators $c_j$ and $c_j^\dagger$, 
\begin{alignat*}{3}
c_j &= \left(\prod_{k=1}^{j-1}(-\sigma^z_k)\right)\sigma^-_j, \qquad  \qquad  \sigma^-_j &= \left(\prod_{k=1}^{j-1}(-\sigma^z_k)\right)c_j, \\
c_j^\dagger &= \left(\prod_{k=1}^{j-1}(-\sigma^z_k)\right)\sigma^+_j, \qquad  \qquad \sigma^+_j &= \left(\prod_{k=1}^{j-1}(-\sigma^z_k)\right)c_j^\dagger, 
\end{alignat*}
which satisfy the usual anti-commutation relations for fermions, 

\begin{equation*}
\{c_j^\dagger, c_{j'} \} = \delta_{j,j'}, \qquad \qquad \{c_j, c_{j'} \} = \{c_j^\dagger, c_{j'}^\dagger \} = 0.
\end{equation*}
The $c_j$ and $c_j^\dagger$ are real matrices and are indeed conjugate to one another. With this transformation, 

\begin{equation*}
H =  \sum_{j=1}^{N-1} \left( c^\dagger_j c_{j+1} + c^\dagger_{j+1} c_{j}  \right) -i (c^\dagger_1 c_1 - c^\dagger_N c_N),
\end{equation*}

\noindent which can also be written as

\begin{equation}
H =  \sum_{k_1,k_2} c^\dagger_{k_1} c_{k_2} \, \, \mathcal{N}_{k_1,k_2},
\label{eq:HN}\end{equation}

\noindent where 

\begin{equation*}
\mathcal{N} = 
		\begin{pmatrix} -i & 1 & 0 & 0 &\dots & 0 & 0 & 0  \\
                  1 & 0 & 1 & 0 &\dots & 0 & 0 & 0\\
                  0 & 1& 0  & 1 &\dots & 0 & 0 & 0\\
                  0& 0 & 1& 0  &\dots & 0 & 0 & 0\\ 
                  \vdots & \vdots& \vdots  &\vdots  & \ddots  & \vdots & \vdots & \vdots\\ 
                   0 & 0  & 0 & 0 &\dots & 0 & 1 & 0\\ 
                   0& 0& 0&0 &\dots &1  & 0& 1\\
                   0 &0 &0 &0 & \dots& 0 & 1 & i\end{pmatrix}
\end{equation*}

\noindent is a symmetric matrix (but not a hermitian matrix) of size $N$. We want to perform a Bogoliubov transformation 
\begin{equation}b_n= \sum_j f^j_n c^\dagger_j, \qquad a_n= \sum_j g^j_n c_j,\label{eq:Bogo}\end{equation} that will make $H$ as simple as possible in terms of these new operators. We also require that the $a_n$s and $b_n$s satisfy the fermionic anticommutation relations
\begin{equation}
\{b_n, a_{n'} \} = \delta_{n,n'}, \qquad \qquad \{b_n, b_{n'} \} = \{a_n, a_{n'} \} = 0.
\label{eq:comm}\end{equation}
To this intent, we want to diagonalize $\mathcal{N}$. Define the matrix $\mathcal{K}_L$, of dimensions $L \times L$
\begin{equation*}
\mathcal{K}_L = 
                  \begin{pmatrix} 0 & 1 & 0 & 0 &\dots & 0 & 0 & 0  \\
                  1 & 0 & 1 & 0 &\dots & 0 & 0 & 0\\
                  0 & 1& 0  & 1 &\dots & 0 & 0 & 0\\
                  0& 0 & 1& 0  &\dots & 0 & 0 & 0\\ 
                  \vdots & \vdots& \vdots  &\vdots  & \ddots  & \vdots & \vdots & \vdots\\ 
                   0 & 0  & 0 & 0 &\dots & 0 & 1 & 0\\ 
                   0& 0& 0&0 &\dots &1  & 0& 1\\
                   0 &0 &0 &0 & \dots& 0 & 1 & 0\end{pmatrix}.
\end{equation*}
Also, let $\tilde{\mathcal{N}} = \mathcal{N} - \xi id_N$ and $\tilde{\mathcal{K}}_L = \mathcal{K}_L - \xi id_L$. The eigenvalues of $\mathcal{N}$ are $\xi$s for which  $\det(\tilde{\mathcal{N}})=0$. Summing over the first and last line, we find

\begin{equation*}
\det(\tilde{\mathcal{N}}) = (\xi^2 + 1) \det(\tilde{\mathcal{K}}_{N-2}) + 2 \xi \,\det(\tilde{\mathcal{K}}_{N-3}) + \det(\tilde{\mathcal{K}}_{N-4})
\end{equation*}
and, similarly,
\begin{equation*}
\det(\tilde{\mathcal{K}}_L) = - \xi \det(\tilde{\mathcal{K}}_{L-1}) - \det(\tilde{\mathcal{K}}_{L-2})
\end{equation*}
with initial conditions $\det(\tilde{\mathcal{K}}_1) = -\xi$ and $\det(\tilde{\mathcal{K}}_2) = \xi^2 - 1$ (or, more simply, $\det(\tilde{\mathcal{K}}_0)=1$). These are Chebyshev polynomials of the second type, with recursion relations
\begin{equation*}
U_{k+1}(x) = 2x U_{k}(x) - U_{k-1}(x)
\end{equation*}
and initial conditions $U_0 = 1$ and $U_1(x)=2x$. They can be written in a simple closed form:
$$U_k(\cos v) = \frac{\sin(k+1) v}{\sin v}.$$

\noindent With $\xi = -2 \cos v$, one finds $\det(\tilde{\mathcal{K}}_L) = \sin(L+1)v / \sin v$ and

\begin{alignat}{2}\det(\tilde{\mathcal{N}}) &= \frac{(4 \cos^2 v + 1) \sin(N-1)v}{\sin v} -  \frac{4\cos v \sin (N-2)v}{\sin v} + \frac{\sin (N-3)v}{\sin v} \nonumber \\ &= \frac{2 \cos v \sin Nv}{\sin v} \label{eq:detN}.
\end{alignat}

\noindent Eigenvalues of $\mathcal N$ satisfy one of the two conditions:
\begin{itemize}
\item $\sin Nv / \sin v = 0 $. Solutions for $\xi$ are $ \xi_n = 2 \cos \pi n/N$ with $n = 1, ..., N-1$. (The minus sign has disappeared because we changed $n \leftrightarrow N-n$.) The values $n=0$ and $n=N$ are absent because of the $\sin v$ in the denominator of (\ref{eq:detN}).
\item $\cos v = 0$, with solution $\xi_{N/2} = 0$ (even when $N$ is not even).
\end{itemize}
When $N$ is odd, $v_n= \pi n/N$ is never $\pi/2$. All eigenvalues are distinct and $\mathcal{N}$ is diagonalizable. When $N$ is even however, the eigenvalue $\xi = 0$ appears twice.
\\
\\
\noindent For a fixed value of $n$ in the interval $1, ..., N-1$, we now look for $u_n = (u_n^1, ..., u_n^N)$, the eigenvector of $\mathcal{N}$ with eigenvalue $\xi_n$. Its components satisfy the constraints
\begin{align*} u_n^j - \xi_n u_n^{j+1} + u_n^{j+2} &= 0, \qquad \textrm{for} \qquad j = 1, ..., N-2, \\
 (-i - \xi_n) u_n^1 + u_n^2 &=0,  \\
 u_n^{N-1} + (i - \xi_n) u_n^{N} &=0. \end{align*}

\noindent Let $x_n$ such that $\xi_n = x_n + x_n^{-1}$ (and $x_n= e^{i \pi n/N} $). One can easily verify that the ansatz
\begin{equation} u_n^{j} = K_n (\alpha_n x_n^j+ \gamma_n x_n^{-j}) \qquad \textrm{with} \qquad \alpha_n = -(1+i x_n^{-1}) \qquad \textrm{and} \qquad \gamma_n = 1+ i x_n \label{eq:ansatz}\end{equation}
satisfies all three constraints. For reasons that will be soon clear, when $n \neq N/2$, we fix the constant $K_n$ to $(2 \alpha_n \gamma_n N)^{-1/2}$ ensuring that $u_n^T u_n = 1$. Indeed, 
\begin{align*}
u_n^T u_n &= \sum_{j=1}^N (u_n^j)^2 = \frac1{2\alpha_n\gamma_n N}\sum_{j=1}^N(\alpha_n x_n^j + \gamma_n x_n^{-j})^2 \\
& = 1 + \frac{\alpha_n^2}{2 \alpha_n \gamma_n N}\frac{x_n^2 (1-x_n^{2N})}{1-x_n^2} + \frac{\gamma_n^2}{2 \alpha_n \gamma_n N}\frac{x_n^{-2} (1-x_n^{-2N})}{1-x_n^{-2}} = 1,
\end{align*} 
because $x_n^{\pm 2N}=1$. Notice that we have
$$\alpha_n \gamma_n = -i (x_n + x_n^{-1}) = -i \xi_n.$$
For the states with $\xi = 0$, the cases $N$ odd and $N$ have to be treated separately.

\subsection{N odd}
\label{sec:diagNimpair}
For the eigenvector with $\xi = 0$, the ansatz (\ref{eq:ansatz}) still works with $x=i$. Then, $\gamma_n = 0$, $\alpha_n = -2$ and we can write $u_{N/2}^j = K'_{N/2} i^j $.
$$u_{N/2}^T u_{N/2} = (K'_{N/2})^2 \sum_{j=1}^N (-1)^j = - (K'_{N/2})^2$$
and $K'_{N/2} = i$ is the correct choice. When $N$ is odd, $\mathcal{N}$ is diagonalizable and from (\ref{eq:Bogo}), $H$ can be written as
$H = \sum_{k=0}^{N-1}{\Lambda_k \, b_k a_k}$,
and
$$[H,a_m] = \sum_{k=0}^{N-1} \Lambda_k [b_ka_k,a_m] = -\Lambda_m a_m, \qquad \{c_i^\dagger,  [H,a_m]\} = -\Lambda_m \sum_j g_m^j \{ c_i^{\dagger}, c_j\} = -\Lambda_m g_m^i,$$
but because of (\ref{eq:HN}), we also have
$$[H,a_m] = \sum_{k_1,k_2} \sum_j \mathcal{N}_{k_1,k_2} \,g_m^j [c_{k_1}^\dagger c_{k_2}, c_j] = - \sum_{k_1,k_2} \mathcal{N}_{k_1, k_2}\, g_m^{k_1} c_{k_2},$$
$$\{c_i^\dagger,  [H,a_m]\} = - \sum_{k_1,k_2} \mathcal{N}_{k_1,k_2}\, g_m^{k_1} \{c_i, c_{k_2}\} = -\sum_{k_1} \mathcal{N}_{i,k_1} g_m^{k_1},$$
where we used $\mathcal{N}_{i,j} = \mathcal{N}_{j,i}$. We can write
\begin{equation}\mathcal{N} \vec{g}_m = \Lambda_m \vec{g}_m, \qquad \textrm{where} \qquad 
\vec{g}_m = \begin{pmatrix} g_m^1 \\ g_m^2 \\ \vdots \\ g_m^{N}\end{pmatrix}.\label{eq:Ng}\end{equation}
The $g^j_m$s are the components of the eigenvectors of $\mathcal{N}$ and the $\Lambda_m$s, its eigenvalues. The same process can be carried out for the $b_m$s, yielding
\begin{equation}\mathcal{N} \vec{f}_m = \Lambda_m \vec{f}_m, \qquad \textrm{where} \qquad 
\vec{f}_m = \begin{pmatrix} f_m^1 \\ f_m^2 \\ \vdots \\ f_m^{N}\end{pmatrix}.\label{eq:Nf}\end{equation}

\noindent The labeling of the $a$s and $b$s is as follows.
\begin{itemize}
\item For $ n = 1, ..., N-1$, we choose $\Lambda_n = \xi_n \, \textrm{and} \, f_n^j=g_n^j =u_n^j  \,  $. This gives
\begin{equation} a_n = K_n \sum_{j=1}^N (\alpha_n x_n^j+ \gamma_n x_n^{-j}) c_j, \qquad \qquad b_n = K_n \sum_{j=1}^N  (\alpha_n x_n^j+ \gamma_n x_n^{-j}) c_j^\dagger, \label{eq:ab}\end{equation}
with $K_n$, $\alpha_n$ and $\gamma_n$ given previously.
\item For the eigenvector with eigenvalue zero, $\Lambda_0 = \xi_{N/2} = 0$, $f_0^j = g_0^j = u^j_{N/2} = i^{j+1}$, and
\begin{equation} 
a_0 = \sum_{j=1}^N i^{j+1} c_j, \qquad \qquad b_0 = \sum_{j=1}^N  i^{j+1} c_j^\dagger.
\label{eq:ab0}\end{equation}
\end{itemize}
Because  $f_k^j$ has a non zero imaginary part and $f_k^j = g_k^j$, $b_n \neq a_n^\dagger$. Instead, $c_j^\dagger = c_j^T$ gives $b_n = a_n^T$. In terms of $f_k^j$ and $g_k^j$, the constraint given by the anticommutation relation is 
$$\delta_{n,n'} = \{ b_n, a_{n'}\} = \sum_{j,j'} g_n^j g_{n'}^{j'} \{c_j^\dagger, c_{j'} \} = \sum_{j} g^j_n g^j_{n'} = \vec{g}^T_n \vec{g}_{n'}.$$
When $n \neq n'$, this is trivial, because
$$0 = \vec{g}^T_n (\mathcal{N}-\mathcal{N}^T)\vec{g}_{n'} = \vec{g}^T_n \vec{g}_{n'} (\xi_{n'} - \xi_{n}) \qquad \textrm{and} \qquad \xi_n \neq \xi_{n'}.$$
However when $n=n'$, $\vec{g}^T_n \vec{g}_{n} = 1$ explains our previous choice for the $K_n$s. Finally, one finds that $H$ can be written as $H = 2 \sum_{k=1}^{N-1} \cos(\pi k/N) b_k a_k$. If we denote by $|0\rangle$ the state $|\uparrow \uparrow \dots \uparrow \, \rangle $ with all spins up, then eigenvectors of $H$ in the sector $S^z = N/2 - n$ are 

\begin{equation}|\gamma\rangle = a_{k_1}a_{k_2} \dots a_{k_n} |0\rangle, \label{eq:eigenvec}\end{equation}

\noindent where the $k_1, ..., k_n$ are in the interval $0, ..., N-1$ and appear at most once. When the $a_0$ excitation is present, we decide to set it at the end, $k_n = 0$. With this convention, the eigenvalue of $|\gamma \rangle$ is
\begin{equation*} \gamma = \left\{ \begin{array}{l l}  2\sum_{i=1}^n  \cos(\pi k_i /N),  & \textrm{if} \,\, k_n \neq 0, \\ 2\sum_{i=1}^{n-1}  \cos(\pi k_i /N),  & \textrm{if} \,\, k_n = 0. \end{array}\right.\end{equation*}

\subsection{N even}
\label{sec:diagNpair}
For $N$ even, the eigenvalue $0$ appears twice and $\mathcal{N}$ is not diagonalizable. To show this, we study $\mathcal{N}^2$:

\begin{equation*}
\mathcal{N}^2 = 
		\left( \begin{array}{c c c c c c c c c c c c c} 
		0 & -i & 1 & 0 & 0 & 0 & \dots & 0 & 0 & 0 & 0 & 0 & 0  \\
                  -i & 2 & 0 & 1 & 0 & 0 & \dots & 0 & 0 & 0 & 0 & 0 & 0  \\
                  1 & 0& 2  & 0 & 1 & 0 & \dots & 0 & 0 & 0 & 0 & 0 & 0  \\
                  0 & 1& 0  & 2  & 0 & 1 & \dots & 0 & 0 & 0 & 0 & 0 & 0  \\
                   0 & 0 & 1  & 0 &2 & 0 & \dots & 0 & 0 & 0 & 0 & 0 &  0 \\
                   0 & 0& 0 & 1& 0 & 2  & \dots & 0 & 0 & 0 & 0 & 0 & 0  \\
		\vdots & \vdots& \vdots & \vdots&\vdots & \vdots & \ddots& \vdots & \vdots & \vdots & \vdots & \vdots & \vdots \\
		0 & 0& 0& 0& 0& 0 & \dots & 2  & 0& 1 & 0 & 0 & 0 \\
		0 & 0& 0& 0& 0& 0& \dots & 0 & 2  & 0& 1 & 0 & 0 \\
		0 & 0& 0& 0& 0& 0& \dots & 1& 0 & 2  & 0& 1 & 0 \\
		0 & 0& 0& 0& 0& 0& \dots & 0& 1 & 0  & 2 & 0& 1 \\
		0 & 0& 0& 0& 0&0 & \dots & 0 & 0& 1 & 0  & 2& i \\                   
		0 & 0& 0& 0& 0& 0& \dots & 0 & 0 & 0 & 1 & i & 0\\
                  \end{array} \right)
\end{equation*}

\noindent and check that
$$ w_1^j = i^j \Big \lfloor \frac{N-j-1}2 \Big\rfloor \qquad \textrm{and} \qquad  w_2^j = i^j \Big\lfloor \frac{N-j+1}2\Big\rfloor $$
are two independent eigenvectors of $\mathcal{N}^2$ with eigenvalue $0$. The eigenvector $u^j_{N/2} = K'_{N/2} i^j$ of $\mathcal{N}$ is given by the linear combination $w_2^j - w_1^j = i^j$  (though the constant $K'_{N/2}$ will be different from the $N$ odd constant). Also,
\begin{equation}(\mathcal{N} w_1)^{j} 
= i^{j-1}, \qquad  (\mathcal{N} w_2)^{j} = i^{j-1},\label{eq:actionNw}\end{equation}
and any linear combination $ w = \beta_1  w_1 + \beta_2  w_2$ satisfies $\mathcal{N} w  \propto u_{N/2}$; $\mathcal{N}$ is therefore not diagonalizable. Nevertheless, it is possible to write $H$ in the following manner:

\begin{equation} H = b_{0}a_{-1} + \displaystyle\sum_{\substack{n = 1\\ n\neq N/2}}^{N-1} \Lambda_n b_n a_n \label{eq:Hnondiag}\end{equation}

\noindent where all the $a$s and $b$s obey (\ref{eq:comm}). The identification for $N$ even is slightly modified:

\begin{itemize}
\item For the $N-2$ eigenvecteurs with $\xi \neq 0$, (\ref{eq:Ng}) and (\ref{eq:Nf}) stay valid and the same identification is made: $\Lambda_n = \xi_n = 2 \cos{\pi n/N}$ and $f^j_n = g^j_n = u^j_n$ (for $n = 1,2,..., N-1$, except $n=N/2$). The operators $a$ and $b$ are then given by the solution (\ref{eq:ab}).
\item For the two remaining modes, a new feature appears:
\begin{equation*}\begin{array}{l l l}
0 = -[H,a_{-1}] = \displaystyle{\sum_{k_1,k_2}} g_{-1}^{k_1}\mathcal{N}_{k_1,k_2}\, c_{k_2}  \qquad & \rightarrow \qquad \mathcal{N}\vec{g}_{-1} &=0,\\ 
a_{-1} = -[H,a_{0}] = \displaystyle{\sum_{k_1,k_2}} g_{0}^{k_1}\mathcal{N}_{k_1,k_2}\, c_{k_2} \qquad & \rightarrow \qquad \mathcal{N}\vec{g}_{0} &=\vec{g}_{-1},\\ 
0 = [H,b_{0}] = \displaystyle{\sum_{k_1,k_2}} f_{0}^{k_1}\mathcal{N}_{k_1,k_2}\, c_{k_2} \qquad & \rightarrow \qquad \mathcal{N}\vec{f}_{0} &=0,\\ 
b_{0} = [H,b_{-1}] = \displaystyle{\sum_{k_1,k_2}} f_{-1}^{k_1}\mathcal{N}_{k_1,k_2}\, c_{k_2}  \qquad & \rightarrow \qquad \mathcal{N}\vec{f}_{-1} &=\vec{f}_0,
  \end{array}\end{equation*}
  where the equations on the right are obtained by anti-commuting the equations on the left with $c_i$ and writing the result as matrix products.
The result is $f_{0}^j = g_{-1}^j = u_{N/2}^j = K'_{N/2} i^j$ and $f_{-1}^j = g_{0}^j = w^j = \beta_1  w_1^j + \beta_2  w_2^j$, where  $K'_{N/2}$, $\beta_1$ and $\beta_2$ are constants that remain to be fixed. The relation $\mathcal{N}w = u_{N/2}$, along with the commutation relations (\ref{eq:comm}), fixes these constants (this is done in appendix \ref{app:a}). The final result is
%
\begin{alignat}{3} 
& a_{0} = \sum_{j=1}^N (\beta_1 w_1^j + \beta_2 w_2^j) c_j, &\qquad \qquad  &b_{0} = K'_{N/2}\sum_{j=1}^N  i^{j} c_j^\dagger, 
\label{eq:ab0pair} \\
& a_{-1} = K'_{N/2}\sum_{j=1}^N i^{j} c_j, &\qquad \qquad  &b_{-1} = \sum_{j=1}^N  (\beta_1 w_1^j + \beta_2 w_2^j) c_j^\dagger, \nonumber
\end{alignat}
with $K'_{N/2} = (2i/N)^{1/2}$, $\beta_1 = \frac{-1}{2K_{N/2}}$ and $\beta_2 = -\frac{N-4}{N} \beta_1$. The new feature here is the pairing $a_0^T = b_{-1}$ and $a_{-1}^T = b_0$. 
\end{itemize}

\noindent Finally, the canonical expression for the Hamiltonian is $$ H = b_{0}a_{-1} + 2 \displaystyle\sum_{\substack{k = 1\\ k\neq N/2}}^{N-1} \cos (\pi k/N) b_k a_k.$$ 

\noindent In the sector $S^z = N/2 - n$, the states $|\gamma \rangle$ given in eq. \hspace{-0.1cm}(\ref{eq:eigenvec}) are tied to the eigenvalues

\begin{equation*} \gamma = \left\{ \begin{array}{l l}  2\sum_{i=1}^n  \cos(\pi k_i /N),  & \textrm{if} \,\, a_0 \,\,\textrm{and} \,\, a_{-1} \,\,\textrm{are absent}, \\ 2\sum_{i=1}^{n-1}  \cos(\pi k_i /N),  & \textrm{if only one of} \,\, a_0 \,\,\textrm{or} \,\, a_{-1} \,\,\textrm{is present,} \\ 2\sum_{i=1}^{n-2}  \cos(\pi k_i /N),  & \textrm{if both} \,\, a_0 \,\,\textrm{and} \,\, a_{-1} \,\,\textrm{are present.}\end{array}\right.\end{equation*}

All the $k_i$s are in the set $\{-1, 0, ..., N-1\} \setminus \{N/2\}$ and, as in the $N$ odd case, the $a_0$ and $a_{-1}$ are always set to the last $k_i$s, when present. Not all the states $|\gamma\rangle$ are eigenstates of $H$. The generalized eigenvectors are those with the $a_{0}$ excitation, but not $a_{-1}$. In total, there are $2^{N-2}$ such states, while all others are eigenvectors.

%
\section{The algebra $U_q(sl_2)$}\label{sec:uqsl2}
%
The algebra $U_q(sl_2)$ is generated by the three generators $q^{S^z}$, $S^+$ and $S^-$ that satisfy the relations
$$ q^{S^z} S^\pm q^{-S^z} = q^{\pm1} S^\pm \qquad \textrm{and} \qquad [S^+,S^-] = \frac{q^{2S^z}-q^{-2S^z}}{q-q^{-1}}.$$

\begin{Proposition}
The representation 
\begin{align*} 
q^{S^z} &= q^{\sigma^z /2} \otimes q^{\sigma^z /2} \otimes \dots \otimes q^{\sigma^z /2} = \prod_{j=1} ^Nq^{\sigma_j^z/2}, \\ 
S^z &= \sum_{j=1}^N \sigma_j^z/2, \\ 
S^\pm &= \sum_{j=1}^N S^\pm_j= \sum_{j=1}^N  q^{-\sigma^z /2} \otimes ... \otimes q^{-\sigma^z /2} \otimes \sigma^\pm \otimes q^{\sigma^z /2}\otimes \dots \otimes q^{\sigma^z /2} \\ & = \sum_{j=1}^{N} \left(\prod_{k=1}^{j-1} q^{-\sigma_k^z/2}\right) \sigma^\pm_j \left(\prod_{k'=j+1}^{N} q^{\sigma_k^z/2}\right)
 \end{align*}
of $U_q(sl_2)$ commutes with the $e_i$ matrices given in (\ref{eq:eTL}). \label{sec:Uqcommutes}
\end{Proposition}
\noindent{\scshape Proof\ \ } The commutation of $q^{S^z}$, $S^+$ and $S^-$ with $e_i$ arises from the relations
 $$[\tilde {e}, q^{\sigma^z/2} \otimes q^{\sigma^z/2}] =0 \qquad \textrm{and} \qquad [\tilde{e}, q^{-\sigma^z/2} \otimes \sigma^\pm + \sigma^\pm \otimes  q^{\sigma^z/2}]=0,$$
 where $\tilde e$ is the $4\times 4$ matrix given in (\ref{eq:etilde}).
\hfill$\square$ 

This property, first noticed in \cite{PasquierSaleur}, will be used thoroughly. Note also that $S^- = (S^+)^T$.
Some particularities occur when $q^{2P} = 1$. Let $q_c$ be a $2P$-th root of unity. Then $(S^\pm)^P|_{q=q_c} =0$. For these values $q_c$, the generators $(S^\pm)^P$ can be replaced by (\cite{Lusztig}, \cite{PasquierSaleur}):
\begin{equation*}
S^{\pm(P)} \equiv \lim_{q \rightarrow q_c} \frac{(S^\pm)^P}{[P]_q!}, \qquad \textrm{where} \qquad {[n]_q!} = \prod_{k=1}^{n} [n]_q \qquad \textrm{and} \qquad [n]_q = \frac{q^n - q^{-n}}{q-q^{-1}}.
\end{equation*}

\noindent For $q = q_c$ a root or unity, $S^{\pm (P)}$ is non zero and commutes with $e_i$, because
\begin{equation*}
[S^{\pm (P)},e_i] = \lim_{q\rightarrow q_c} \frac{[(S^{\pm})^P,e_i]}{[P]_q} = \lim_{q\rightarrow q_c} \frac{0}{[P]_q} =0.
\end{equation*}

\noindent We are interested in the case $q_c=i, P=2$,  and calculate $S^{\pm(2)}$. The square of $S^\pm$ is
\begin{equation*}
(S^\pm)^2 = \sum_{j_1,j_2} S_{j_1}^\pm S_{j_2}^\pm = \left(\sum_{j_1<j_2} +\sum_{j_2<j_1}  \right) S_{j_1}^\pm S_{j_2}^\pm = \sum_{j_1<j_2}(S_{j_1}^\pm S_{j_2}^\pm + S_{j_2}^\pm S_{j_1}^\pm).
\end{equation*}
When ${j_1<j_2}$,
\begin{align*}
S_{j_1}^\pm S_{j_2}^\pm &= \left( \prod_{k_1=1}^{j_1-1} q^{-\sigma^z_{k_1}} \right) \sigma_{j_1}^\pm q^{-\sigma_{j_1}^z/2}  q^{\sigma_{j_2}^z/2} \sigma_{j_2}^\pm  \left( \prod_{k_2=j_2+1}^{N} q^{\sigma^z_{k_2}} \right) \\
&= q^{\pm 1}  \left( \prod_{k_1=1}^{j_1-1} q^{-\sigma^z_{k_1}} \right) \sigma_{j_1}^\pm  \sigma_{j_2}^\pm  \left( \prod_{k_2=j_2+1}^{N} q^{\sigma^z_{k_2}} \right),
\end{align*}
but
\begin{equation*}
 S_{j_2}^\pm S_{j_1}^\pm = q^{\mp 1}  \left( \prod_{k_1=1}^{j_1-1} q^{-\sigma^z_{k_1}} \right) \sigma_{j_1}^\pm  \sigma_{j_2}^\pm  \left( \prod_{k_2=j_2+1}^{N} q^{\sigma^z_{k_2}} \right)
\end{equation*}
and finally,
\begin{equation*}
\frac{(S^{\pm})^2}{[2]_q} = \sum_{j_1<j_2} \underbrace{\left( \prod_{k_1=1}^{j_1-1} q^{-\sigma^z_{k_1}} \right) \sigma_{j_1}^\pm  \sigma_{j_2}^\pm  \left( \prod_{k_2=j_2+1}^{N} q^{\sigma^z_{k_2}} \right)}_{S^{\pm(2)}_{j_1,j_2}(q)}.
\end{equation*}
%

\subsection{$S^\pm$ and $S^{\pm(2)}$ for free fermions}
The next step is to write $S^\pm$ and $S^{\pm(2)}$ first in terms of operators $c_j$ and $c^\dagger_j$, and then of the $a_n$s and $b_n$s calculated in section \ref{sec:XXZ} (Deguchi {\it et al.} did this for the periodic case \cite{Deguchi}). We start with $S^+$ and $S^-$,
\begin{align*}
S^\pm &= \left(\prod_{k=1}^{N} q^{\sigma_k^z/2}\right)  \sum_{j=1}^{N} \left(\prod_{k=1}^{j-1} q^{-\sigma_k^z}\right) q^{-\sigma^z_j/2} \sigma^\pm_j \\
 &= q^{S^z} q^{\mp1/2} \sum_{j=1}^{N} \left(\prod_{k=1}^{j-1} q^{-\sigma_k^z}\right) \sigma^\pm_j \\
 &= i^{S^z \mp1/2} \sum_{j=1}^{N} \left(\prod_{k=1}^{j-1} -i \sigma_k^z \right) \sigma^\pm_j \\
 &= i^{S^z \mp1/2 -1} \sum_{j=1}^{N} i^{j}\left(\prod_{k=1}^{j-1} - \sigma_k^z \right) \sigma^\pm_j \\
\end{align*}
and this yields
\begin{equation}
S^+ = i^{S^z -3/2} \sum_{j=1}^{N} i^{j}c_j^\dagger = \frac{i^{S^z - 3/2}}{K'_{N/2}} \sum_{j=1}^{N} u^j_{N/2}c_j^\dagger \qquad \textrm{and} \qquad S^- =  i^{S^z -1/2} \sum_{j=1}^{N} i^{j}c_j = \frac{i^{S^z -1/2}}{K'_{N/2}} \sum_{j=1}^{N} u^j_{N/2}c_j.
\label{eq:S+S-c}
\end{equation}
\noindent We can repeat the computation for $S^{+(2)}$ and $S^{-(2)}$:
\begin{equation*}
S^{+(2)} = i^{-1}(-1)^{S^z} \sum_{j_1<j_2} i^{j_1+j_2} c_{j_1}^\dagger c_{j_2}^\dagger \qquad \textrm{and}\qquad S^{-(2)} = -i^{-1}(-1)^{S^z} \sum_{j_1<j_2} i^{j_1+j_2} c_{j_1} c_{j_2}.
\end{equation*}
Though it is less apparent than before, both $S^- = (S^+)^T$ and $S^{-(2)} = (S^{+(2)})^T$ still hold. Our ultimate goal is to write $S^{+(2)}$ and $S^{-(2)}$ as
\begin{equation*}
S^{+(2)} = \sum_{k_1,k_2} A(k_1, k_2) b_{k_1} b_{k_2},  \qquad
S^{-(2)} = -\sum_{k_1,k_2} A(k_1, k_2) b^T_{k_1} b^T_{k_2},
\end{equation*}
where $A(k_1, k_2) = - A(k_2, k_1)$. To do this calculation, we need to find the inverse formula
\begin{equation*}
c_j^\dagger= \sum_k d^k_j b_k, \qquad \qquad  c_j= \sum_k e^k_j a_k.
\end{equation*}
To do so, we calculate $\{ c_j^\dagger, a_k\}$ and $\{ c_j, b_k\}$ in the two possible ways, to find $d^k_j = e^k_j =g_k^j$. This allows us to pursue the computation,
\begin{align*}
S^{+(2)} &= \frac{i^{-1}(-1)^{S^z}}{2} \sum_{j_1<j_2} i^{j_1+j_2} (c_{j_1}^\dagger c_{j_2}^\dagger-c_{j_2}^\dagger c_{j_1}^\dagger)  \\
& = \frac{i^{-1}(-1)^{S^z}}{2} \sum_{k_1,k_2} b_{k_1} b_{k_2}  \underbrace{\left(\sum_{j_1<j_2} i^{j_1+j_2} (g_{k_1}^{j_1}g_{k_2}^{j_2}-g_{k_1}^{j_2}g_{k_2}^{j_1})\right)}_{B(k_1,k_2)}
\end{align*}
and $B(k_1,k_2)$ can be calculated directly. For any $k_1, k_2$ with $\xi \neq 0$,
\begin{alignat*}{3}
B(k_1,k_2) &= K_{k_1} K_{k_2} \sum_{j_1<j_2} i^{j_1+j_2} \Big(  \alpha_{k_1}  \alpha_{k_2} (x_{k_1}^{j_1}x_{k_2}^{j_2}-x_{k_1}^{j_2}x_{k_2}^{j_1}) + \gamma_{k_1}\gamma_{k_2} (x_{k_1}^{-j_1}x_{k_2}^{-j_2}-x_{k_1}^{-j_2}x_{k_2}^{-j_1})   \\
& \qquad \qquad \qquad \qquad \qquad+\alpha_{k_1}\gamma_{k_2} (x_{k_1}^{j_1}x_{k_2}^{-j_2}-x_{k_1}^{j_2}x_{k_2}^{-j_1}) 
+\alpha_{k_2}\gamma_{k_1} (x_{k_1}^{-j_1}x_{k_2}^{j_2}-x_{k_1}^{-j_2}x_{k_2}^{j_1}) \Big) \\
&=  K_{k_1} K_{k_2} \left(  g(x_{k_1},x_{k_2}) + \, g(x_{k_1}^{-1},x_{k_2}^{-1}) - g(x_{k_1},x^{-1}_{k_2}) - g(x^{-1}_{k_1},x_{k_2})\right)
\end{alignat*}

\noindent where $ g(z,w) = (f(z,w)- f(w,z))(1+iz^{-1})(1+iw^{-1})$  and $ f(z,w) = \sum_{j_1<j_2} (iz)^{j_1} (iw)^{j_2}$. After simplification, one finds
$$g(z,w) = \big((iz)^N-(iw)^N \big) +\frac{  (iw -iz) \big(1-(-zw)^N \big)}{1+zw}$$
and
\begin{equation}B(k_1,k_2) = \frac{i (-1)^{N}K_{k_1}K_{k_2}}{(x_{k_2}+x_{k_1})(1+x_{k_2}x_{k_1})(x_{k_2}x_{k_1})^N}\left( (x_{k_1}^{2N}-x_{k_2}^{2N})(1-x_{k_1}^2x_{k_2}^2) + (1-x_{k_2}^{2N}x_{k_1}^{2N})(x_{k_2}^2-x_{k_1}^2)\right)\label{eq:bk1k2}\end{equation}
Because $x_{k_i} = e^{i \pi k_i/N}$, $x_{k_i}^{2N}=1$ and $B(k_1,k_2)=0$ in general. There is an exception when $x_{k_1}x_{k_2} = -1$, i.e.  \hspace{-0.1 cm}when $k_1+k_2 = N$. $B(k_1,N-k_1)$ is calculated by taking the limit
\begin{equation*}B(k_1,N-k_1) = \lim_{x_{k_2}\rightarrow -1/x_{k_1}} B(k_1,k_2). \end{equation*}
The first term is zero, but not the second,
\begin{equation*}
B(k,N-k) = -2 N i K_{k} K_{N-k}(x_k+x_k^{-1}) = -2N i K_{k} K_{N-k} \xi_k.
\end{equation*}
This simplifies even more, because when $k<N/2$, $$K_kK_{N-k} = \frac1{2N (\alpha_k \gamma_k \alpha_{N-k}\gamma_{N-k})^{1/2}} = \frac{1}{2N (-\xi_k \xi_{N-k})^{1/2}} = \frac1{2N\xi_k},$$
and finally, 
\begin{equation}
B(k_1,k_2) = 
 \left\{ 
\begin{array}{l l}
  -i \delta_{k_1+k_2,N} & \quad  k_1<N/2, \\
  i \delta_{k_1+k_2,N} & \quad  k_1>N/2. \end{array} \right.
\label{eq:Bf}
\end{equation}

\subsection{$N$ odd}
From (\ref{eq:ab0}) and (\ref{eq:S+S-c}), one finds directly 
\begin{equation} S^+ = i^{S^z + 3/2} b_0 \quad \textrm{and} \quad S^- = i^{S^z - 3/2} a_0.\label{eq:Sodd}\end{equation} For $S^{+(2)}$,  $B(k_1,k_2)$ has been calculated except when $k_1 = 0$. 
The result (\ref{eq:bk1k2}) for $B(k_1,k_2)$ is also valid for $k_1= 0$ (as the eigenstate is still given by (\ref{eq:ansatz})); replacing $x_{k_1} =i$ gives $B(0,k) =0$ for all values of $k$ in $1,...,N-1$, and
\begin{align} S^{+(2)} &= \frac{(-1)^{S^z+1}}{2} \left(\sum_{k=1}^{(N-1)/2} b_{k}b_{N-k} - \sum_{k=(N+1)/2}^{N-1} b_{k}b_{N-k}\right) = (-1)^{S^z+1} \sum_{k=1}^{(N-1)/2} b_{k}b_{N-k}, \label{eq:S+-odd}\\
S^{-(2)} &= \frac{(-1)^{S^z}}{2} \left(\sum_{k=1}^{(N-1)/2} a_{k}a_{N-k} - \sum_{k=(N+1)/2}^{N-1} a_{k}a_{N-k}\right) = (-1)^{S^z} \sum_{k=1}^{(N-1)/2} a_{k}a_{N-k} \nonumber.
\end{align}
Because the operators $b_{k}b_{N-k}$ and $a_{k}a_{N-k}$ commute with $H$, $S^{+(2)}$ and $S^{-(2)}$ also do, as expected.
\subsection{$N$ even}
The case $N$ even is again different because of the Jordan cell of size $2$ in $\mathcal{N}$ related to the eigenvalue $0$. From (\ref{eq:ab0pair}) and (\ref{eq:S+S-c}),
\begin{equation} S^+= \frac{i^{S^z-3/2}}{K'_{N/2}}b_{0}, \qquad \qquad S^-= \frac{i^{S^z-1/2}}{K'_{N/2}}a_{-1}. \label{eq:Seven}\end{equation}
For $S^{+(2)}$ and $S^{-(2)}$,  $B(k_1,k_2)$ has been calculated for $k_1, k_2$ in $\{1, ..., N-1\} \setminus \{N/2\}$ in (\ref{eq:Bf}). When $k =0$ or $-1$ and $k' \in \{1, ..., N-1\} \setminus \{N/2\}$, as before we can show that $B(0,k') = B(-1,k') =0$. A quick argument consists in noticing that operators $b_0 b_{k'}$ and $b_{-1} b_{k'}$ do not commute with $H$ and that $S^{+(2)}$ could not have a component along these operators. But there is a component $b_{0}b_{-1}$:

\begin{alignat*}{3}
B(0,-1) = K'_{N/2}\beta_1 \sum_{j_1<j_2} (-1)^{j_1+j_2} \Big( \Big\lfloor \frac{N-j_1-1}{2} &\Big\rfloor - \frac{N-4}N \Big\lfloor \frac{N-j_1+1}{2} \Big\rfloor \\ 
 - & \Big\lfloor \frac{N-j_2-1}{2} \Big\rfloor + \frac{N-4}N \Big\lfloor \frac{N-j_2+1}{2} \Big\rfloor \Big).
\end{alignat*}
To evaluate these sums (for $N$ even), note that
\begin{align*}\sum_{j_1<j_2} (-1)^{j_1+j_2}  \Big\lfloor \frac{N-j_1-1}{2} \Big\rfloor &= \sum_{j_1=1}^N (-1)^{j_1}  \Big\lfloor \frac{N-j_1-1}{2} \Big\rfloor \sum_{j_2=j_1+1}^N (-1)^{j_2} \\
&=\sum_{j_1=1}^N (-1)^{j_1}  \Big\lfloor \frac{N-j_1-1}{2} \Big\rfloor \delta_{1,j_1\textrm{mod}2} \\
&= -\sum_{j=1}^{N/2}  (N/2 - j  ) = -\frac{N(N-2)}8
\end{align*}
and in a similar fashion,
\begin{align*}
\sum_{j_1<j_2} (-1)^{j_1+j_2}  \Big\lfloor \frac{N-j_1+1}{2} \Big\rfloor &= -\frac{N(N-2)}8 - \frac{N}2, \\
\sum_{j_1<j_2} (-1)^{j_1+j_2}  \Big\lfloor \frac{N-j_2-1}{2} \Big\rfloor &= -\frac{N(N-2)}8 + \frac{N}2,\\
\sum_{j_1<j_2} (-1)^{j_1+j_2}  \Big\lfloor \frac{N-j_2+1}{2} \Big\rfloor &= -\frac{N(N-2)}8. \\
\end{align*}
After simplification, $B(0,-1) = -2 K'_{N/2}\beta_1 = 1$, and
\begin{align} S^{+(2)} &=  \frac{(-1)^{S^z}}{2} \left( i (b_{-1} b_{0} -b_{0}b_{-1}) - \sum_{k=1}^{(N-2)/2} b_{k}b_{N-k} + \sum_{k=(N+2)/2}^{N-1} b_{k}b_{N-k} \right) \nonumber \\ &= (-1)^{S^z} \left( i b_{-1} b_{0} - \sum_{k=1}^{(N-2)/2} b_{k}b_{N-k} \right), \label{eq:S+-even}\\
S^{-(2)} &= (-1)^{S^z} \left(i a_{-1} a_{0}  + \sum_{k=1}^{(N-1)/2} a_{k}a_{N-k}  \right) \nonumber.
\end{align}

%
\section{The relation between $H$ and $\mathcal{H}_N$}\label{sec:homo}
%

In this section, we make explicit the relation between the $XXZ$ model and the loop model. The results in this section hold for all $q$.
\subsection{The homomorphism}
We start by introducing a notation for link states. Let $v$ be a link state in $ B_N^d$ with $n = (N-d)/2$ bubbles and let $\psi(v) = \{ (p_1, q_1), (p_2, q_2), ..., (p_n, q_n)\}$, where the $p_i$s are the positions where the bubbles of $v$ start and $q_i$s the positions where they end. In $\psi(v)$, the $(p_i, q_i)$ pairs are ordered in ascending order of $p_i$, though this choice will play no role.
\begin{Definition} 
The linear transformation $i_N^d: V_N^d \rightarrow (\mathbb{C}^2)^{\otimes N}|_{S^z = d/2} $ (the subset of $(\mathbb{C}^2)^{\otimes N}$ of spin configurations with $n = (N-d)/2$ down spins) is defined by its action on the basis elements of $B_N^d$,
\begin{equation}
i_N^d(v) = \left( \prod_{(i,j) \in \psi(v)} T_{i,j} \right) | 0 \rangle, \qquad \textrm{where} \qquad T_{i,j} = w \sigma^-_j + w^{-1} \sigma^-_i, 
\label{eq:hom}
\end{equation}
$w= \sqrt{-q}$ and $ | 0 \rangle =  \left(\begin{smallmatrix} 1 \\ 0 \end{smallmatrix}\right) \otimes  \left(\begin{smallmatrix} 1 \\ 0 \end{smallmatrix}\right) \otimes \dots \otimes  \left(\begin{smallmatrix} 1 \\ 0 \end{smallmatrix}\right) = | \uparrow \uparrow \dots \uparrow \, \rangle$ as before. 
\end{Definition}
This definition can seem complex, but its graphical interpretation is not. In the simplest cases,

\begin{equation} i_2^0 \left(
\psset{unit=0.6}
\begin{pspicture}(-0.2,0)(1.2,1)
\psset{linewidth=1pt}
\psdots(0,0)(1,0)
\psset{linecolor=myc2}
\psarc{-}(0.5,0){0.5}{0}{180}
\end{pspicture} \right) = w \, |\uparrow \downarrow \,\rangle + w^{-1} | \downarrow \uparrow \,\rangle, \qquad \qquad 
i_1^1 \left(
\psset{unit=0.6}
\begin{pspicture}(-0.2,0)(0.2,1)
\psset{linewidth=1pt}
\psdots(0,0)
\psset{linecolor=myc2}
\psline{-}(0,0)(0,1)
\end{pspicture} \right) = | \uparrow \, \rangle,
\label{eq:homoexample}
\end{equation}
and when a link state $v$ has more than one bubble or more than one defect, they are replaced recursively by the rule (\ref{eq:homoexample}). For instance,
\begin{align*} i_6^2 \left(
\psset{unit=0.6}
\begin{pspicture}(-0.2,0)(5.2,1)
\psset{linewidth=1pt}
\psdots(0,0)(1,0)(2,0)(3,0)(4,0)(5,0)
\psset{linecolor=myc2}
\psarc{-}(1.5,0){0.5}{0}{180}
\psarc{-}(4.5,0){0.5}{0}{180}
\psline{-}(0,0)(0,1)
\psline{-}(3,0)(3,1)
\end{pspicture} \right) &= 
w^2 \, |\uparrow \uparrow \downarrow \uparrow \uparrow \downarrow \,\rangle +
 \, |\uparrow  \downarrow \uparrow \uparrow \uparrow \downarrow \,\rangle +
 \, |\uparrow \uparrow \downarrow \uparrow \downarrow \uparrow  \,\rangle +
w^{-2} \, |\uparrow  \downarrow \uparrow \uparrow  \downarrow \uparrow \,\rangle , \\
i_6^2 \left(
\psset{unit=0.6}
\begin{pspicture}(-0.2,0)(5.2,1)
\psset{linewidth=1pt}
\psdots(0,0)(1,0)(2,0)(3,0)(4,0)(5,0)
\psset{linecolor=myc2}
\psarc{-}(2.5,0){0.5}{0}{180}
\psline{-}(0,0)(0,1)
\psbezier{-}(1,0)(1,1.4)(4,1.4)(4,0)
\psline{-}(5,0)(5,1)
\end{pspicture} \right) &= 
w^2 \, |\uparrow \uparrow  \uparrow  \downarrow \downarrow \uparrow \,\rangle +
 \,  |\uparrow \uparrow    \downarrow \uparrow \downarrow \uparrow \,\rangle +
 \,  |\uparrow \downarrow  \uparrow  \downarrow \uparrow \uparrow \,\rangle +
w^{-2} \,  |\uparrow   \downarrow \downarrow \uparrow  \uparrow \uparrow \,\rangle.
\end{align*}
The order of pairs $(i,j)$ in $\psi(v)$ is unimportant, as indices in the product (\ref{eq:hom}) are never repeated and $[T_{i,j},T_{k,l}] =0$ when $i,j,k,l$ are all different.

\begin{Proposition}
For any $c \in TL_N(-(q+q^{-1}))$ and any $v \in V_N^d$, $i_N^d(c v |_{d}) = X(c) i_N^d(v)$ where $X(c)$ is the matrix of $c$ in the representation on $(\mathbb{C}^2) ^{\otimes N}$ as given in (\ref{eq:eTL}),  and where $|_d$ means that all components with less than $d$ defects are set to $0$.
\label{sec:fermeture}
\end{Proposition}
\noindent{\scshape Proof\ \ }  To prove the proposition, one must show that the action of the matrix $e_i$ on $i_N^d(v)$ is the same as the action of the generators $U_i$ on link states, except that annihilated defects always give $0$. (We can restrict to the $U_i$s and $e_i$s only, as other connectivities are products of these.) More precisely, let $Y(v) =  \prod_{(m,n)\in \psi'(v)} T_{m,n}$ and $\psi'(v)$ be the subset of $\psi(v)$ that only contains positions of bubbles in $v$ that do not touch $i$, $i+1$, $j$ and $k$. We first give a list of properties sufficient to prove $i_N^d(cv|_d) = X(c)i_N^d(v)$ for any $v$. For each entry of the list, we give a diagrammatic property followed by the algebraic identity that needs to be checked.

\begin{itemize}
\item[1)]
$ X\hspace{-0.1cm}\left(
\psset{unit=0.4}
\begin{pspicture}(-0.2,-1)(1.2,0.4)
\psset{linewidth=1pt}
\rput(0,-2.1){$i$}
\psdots[{linewidth=0.7pt}](0,0)(1,0)
\psdots[{linewidth=0.7pt}](0,-1.5)(1,-1.5)
\psset{linecolor=myc2}
\psarc{-}(0.5,0){0.5}{180}{360}
\psarc{-}(0.5,-1.5){0.5}{0}{180}
\end{pspicture} \right)
 i_N^d \hspace{-0.1cm}\left(
 \psset{unit=0.4}
 \begin{pspicture}(-0.2,0.0)(1.2,0.8)
\psset{linewidth=1pt}
\rput(0,-0.6){$i$}
\psdots[{linewidth=0.7pt}](0,0)(1,0)
\psset{linecolor=myc2}
\psline{-}(0,1)(0,0)
\psline{-}(1,1)(1,0)
\end{pspicture}
\right)
 = X\hspace{-0.1cm} \left(
\psset{unit=0.4}
\begin{pspicture}(-0.2,-1)(2.0,0.4)
\psset{linewidth=1pt}
\rput(0,-2.1){$i$}
\psdots[{linewidth=0.7pt}](0,0)(1,0)
\psdots[{linewidth=0.7pt}](0,-1.5)(1,-1.5)
\psset{linecolor=myc2}
\psline{-}(0,1)(0,0)
\psline{-}(1,1)(1,0)
\psarc{-}(0.5,0){0.5}{180}{360}
\psarc{-}(0.5,-1.5){0.5}{0}{180}
\rput(1.6,-1.5){$|_d$}
\end{pspicture}
  \right) =0 \,\rightarrow \,  e_i  Y(v) |0\rangle = 0, $
 \\
 \item[2)]
$ X\hspace{-0.1cm}\left(
\psset{unit=0.4}
\begin{pspicture}(-0.2,-1)(1.2,0.4)
\psset{linewidth=1pt}
\rput(0,-2.1){$i$}
\psdots[{linewidth=0.7pt}](0,0)(1,0)
\psdots[{linewidth=0.7pt}](0,-1.5)(1,-1.5)
\psset{linecolor=myc2}
\psarc{-}(0.5,0){0.5}{180}{360}
\psarc{-}(0.5,-1.5){0.5}{0}{180}
\end{pspicture} \right)
 i_N^d \hspace{-0.1cm}\left(
 \psset{unit=0.4}
 \begin{pspicture}(-0.2,0.0)(1.2,0.8)
\psset{linewidth=1pt}
\rput(0,-0.6){$i$}
\psdots[{linewidth=0.7pt}](0,0)(1,0)
\psset{linecolor=myc2}
\psarc{-}(0.5,0){0.5}{0}{180}
\end{pspicture}
\right)
  = - (q+q^{-1}) 
   i_N^d \hspace{-0.1cm}\left(
 \psset{unit=0.4}
 \begin{pspicture}(-0.2,0.0)(1.2,0.8)
\psset{linewidth=1pt}
\rput(0,-0.6){$i$}
\psdots[{linewidth=0.7pt}](0,0)(1,0)
\psset{linecolor=myc2}
\psarc{-}(0.5,0){0.5}{0}{180}
\end{pspicture}
\right)
  \, \rightarrow \, e_i  T_{i,i+1} Y(v) |0\rangle = -(q+q^{-1}) T_{i,i+1} Y(v) |0\rangle, $
 \\
\item[3)]
$
X\hspace{-0.1cm}\left(
\psset{unit=0.4}
\begin{pspicture}(-0.2,-1.2)(3.4,0.6)
\psset{linewidth=1pt}
\rput(0,-2.6){$i$}
\rput(3,-2.6){$j$}
\rput(2.1,0){$\dots$}
\rput(2.1,-2){$\dots$}
\psdots[{linewidth=0.7pt}](0,0)(1,0)(3,0)
\psdots[{linewidth=0.7pt}](0,-2)(1,-2)(3,-2)
\psset{linecolor=myc2}
\psline{-}(3,0)(3,-2)
\psarc{-}(0.5,0){0.5}{180}{360}
\psarc{-}(0.5,-2){0.5}{0}{180}
\end{pspicture} 
\right)
 i_N^d \hspace{-0.1cm}\left(
 \psset{unit=0.4}  
 \begin{pspicture}(-0.4,0)(3.2,0.8)
 \rput(2.1,0){$\dots$}
 \rput(0,-0.6){$i$}
\rput(3,-0.6){$j$}
 \psdots[{linewidth=0.7pt}](0,0)(1,0)(3,0)
 \psset{linecolor=myc2}
 \psbezier{-}(1,0)(1,1.2)(3,1.2)(3,0)
 \psline{-}(0,0)(0,1)
  \end{pspicture}\right)
  =  i_N^d \hspace{-0.1cm}\left(
\psset{unit=0.4}  
\begin{pspicture}(-0.4,0)(3.2,0.8)
\psset{linewidth=1pt}
\rput(0,-0.6){$i$}
\rput(3,-0.6){$j$}
\psdots[{linewidth=0.7pt}](0,0)(1,0)(3,0)
\rput(2,0){$\dots$}
\psset{linecolor=myc2}
\psline{-}(3,0)(3,1)
\psarc{-}(0.5,0){0.5}{0}{180}
\end{pspicture} \right)   \rightarrow \, e_i  T_{i+1,j} Y(v) |0\rangle =   T_{i,i+1} Y(v) |0\rangle, 
$
\\
\item[4)]
$
X\hspace{-0.1cm}\left(
\psset{unit=0.4}
\begin{pspicture}(-0.2,-1.2)(3.4,0.6)
\psset{linewidth=1pt}
\psset{linewidth=1pt}
\rput(2,-2.6){$i$}
\rput(0,-2.6){$j$}
\rput(1.1,0){$\dots$}
\rput(1.1,-2){$\dots$}
\psdots[{linewidth=0.7pt}](0,0)(2,0)(3,0)
\psdots[{linewidth=0.7pt}](0,-2)(2,-2)(3,-2)
\psset{linecolor=myc2}
\psline{-}(0,0)(0,-2)
\psarc{-}(2.5,0){0.5}{180}{360}
\psarc{-}(2.5,-2){0.5}{0}{180}
\end{pspicture} 
\right)
 i_N^d \hspace{-0.1cm}\left(
 \psset{unit=0.4}  
 \begin{pspicture}(-0.4,0)(3.2,0.8)
 \rput(2,-0.6){$i$}
 \rput(0,-0.6){$j$}
 \rput(1.1,0){$\dots$}
 \psdots[{linewidth=0.7pt}](0,0)(2,0)(3,0)
 \psset{linecolor=myc2}
\psline{-}(3,0)(3,1)
\psbezier{-}(0,0)(0,1.2)(2,1.2)(2,0)
  \end{pspicture}\right)
  =  i_N^d \hspace{-0.1cm}\left(
 \psset{unit=0.4}  
 \begin{pspicture}(-0.4,0)(3.2,0.8)
 \rput(2,-0.6){$i$}
 \rput(0,-0.6){$j$}
 \rput(1.1,0){$\dots$}
 \psdots[{linewidth=0.7pt}](0,0)(2,0)(3,0)
 \psset{linecolor=myc2}
\psline{-}(0,0)(0,1)
\psarc{-}(2.5,0){0.5}{0}{180}
  \end{pspicture}\right)  \rightarrow e_i  T_{j,i} Y(v) |0\rangle =  T_{i,i+1} Y(v) |0\rangle, 
$
\\
\item[5)]
$
X\hspace{-0.1cm}\left(
\psset{unit=0.4}
\begin{pspicture}(0.8,-1.2)(6.4,0.6)
\psset{linewidth=1pt}
\psset{linewidth=1pt}
\rput(3.1,0){$\dots$}
\rput(3.1,-2){$\dots$}
\rput(5.1,0){$\dots$}
\rput(5.1,-2){$\dots$}
\rput(1,-2.6){$i$}
\rput(4,-2.6){$j$}
\rput(6,-2.6){$k$}
\psdots[{linewidth=0.7pt}](1,0)(2,0)(4,0)(6,0)
\psdots[{linewidth=0.7pt}](1,-2)(2,-2)(4,-2)(6,-2)
\psset{linecolor=myc2}
\psline{-}(4,0)(4,-2)
\psline{-}(6,0)(6,-2)
\psarc{-}(1.5,0){0.5}{180}{360}
\psarc{-}(1.5,-2){0.5}{0}{180}
\end{pspicture} 
\right)
 i_N^d \hspace{-0.1cm}\left(
 \psset{unit=0.4}  
 \begin{pspicture}(0.6,0)(6.4,0.8)
\rput(1,-0.6){$i$}
\rput(4,-0.6){$j$}
\rput(6,-0.6){$k$}
\rput(3.1,0){$\dots$}
\rput(5.1,0){$\dots$}
 \psdots[{linewidth=0.7pt}](1,0)(2,0)(4,0)(6,0)
 \psset{linecolor=myc2}
\psbezier{-}(2,0)(2,1.2)(4,1.2)(4,0)
\psbezier{-}(1,0)(1,2)(6,2)(6,0)
  \end{pspicture}\right)
  =  i_N^d \hspace{-0.1cm}\left(
 \psset{unit=0.4}  
 \begin{pspicture}(0.6,0)(6.4,0.8)
\rput(1,-0.6){$i$}
\rput(4,-0.6){$j$}
\rput(6,-0.6){$k$}
\rput(3.1,0){$\dots$}
\rput(5.1,0){$\dots$}
 \psdots[{linewidth=0.7pt}](1,0)(2,0)(4,0)(6,0)
 \psset{linecolor=myc2}
\psarc{-}(1.5,0){0.5}{0}{180}
\psbezier{-}(4,0)(4,1.2)(6,1.2)(6,0)
  \end{pspicture}\right)
     \rightarrow e_i T_{i,k} T_{i+1,j} Y(v) |0\rangle = T_{i,i+1} T_{j,k} Y(v) |0\rangle, 
$ 
\\
\item[6)]
$
X\hspace{-0.1cm}\left(
\psset{unit=0.4}
\begin{pspicture}(-0.2,-1.2)(5.4,0.6)
\psset{linewidth=1pt}
\psset{linewidth=1pt}
\rput(1.1,0){$\dots$}
\rput(1.1,-2){$\dots$}
\rput(4.1,0){$\dots$}
\rput(4.1,-2){$\dots$}
\rput(2,-2.6){$i$}
\rput(0,-2.6){$j$}
\rput(5,-2.6){$k$}
\psdots[{linewidth=0.7pt}](0,0)(2,0)(3,0)(5,0)
\psdots[{linewidth=0.7pt}](0,-2)(2,-2)(3,-2)(5,-2)
\psset{linecolor=myc2}
\psline{-}(0,0)(0,-2)
\psline{-}(5,0)(5,-2)
\psarc{-}(2.5,0){0.5}{180}{360}
\psarc{-}(2.5,-2){0.5}{0}{180}
\end{pspicture} 
\right)
 i_N^d \hspace{-0.1cm}\left(
 \psset{unit=0.4}  
 \begin{pspicture}(-0.4,0)(5.4,0.8)
\rput(1.1,0){$\dots$}
\rput(4.1,0){$\dots$}
\rput(2,-0.6){$i$}
\rput(0,-0.6){$j$}
\rput(5,-0.6){$k$}
\psdots[{linewidth=0.7pt}](0,0)(2,0)(3,0)(5,0)
 \psset{linecolor=myc2}
\psbezier{-}(0,0)(0,1.4)(2,1.4)(2,0)
\psbezier{-}(3,0)(3,1.2)(5,1.2)(5,0)  \end{pspicture}\right)
  =  i_N^d \hspace{-0.1cm}\left(
 \psset{unit=0.4}  
 \begin{pspicture}(-0.4,0)(5.4,0.8)
\rput(1.1,0){$\dots$}
\rput(4.1,0){$\dots$}
\rput(2,-0.6){$i$}
\rput(0,-0.6){$j$}
\rput(5,-0.6){$k$}
\psdots[{linewidth=0.7pt}](0,0)(2,0)(3,0)(5,0)
 \psset{linecolor=myc2}
\psarc{-}(2.5,0){0.5}{0}{180}
\psbezier{-}(0,0)(0,1.6)(5,1.6)(5,0)  \end{pspicture}\right)
  \rightarrow  e_i T_{j,i} T_{i+1,k} Y(v) |0\rangle = T_{i,i+1} T_{j,k} Y(v) |0\rangle, 
$ 
\\
\item[7)]
$
X\hspace{-0.1cm}\left(
\psset{unit=0.4}
\begin{pspicture}(-0.2,-1.2)(5.4,0.6)
\psset{linewidth=1pt}
\psset{linewidth=1pt}
\rput(1.1,0){$\dots$}
\rput(1.1,-2){$\dots$}
\rput(3.1,0){$\dots$}
\rput(3.1,-2){$\dots$}
\rput(4,-2.6){$i$}
\rput(0,-2.6){$j$}
\rput(2,-2.6){$k$}
\psdots[{linewidth=0.7pt}](0,0)(2,0)(4,0)(5,0)
\psdots[{linewidth=0.7pt}](0,-2)(2,-2)(4,-2)(5,-2)
\psset{linecolor=myc2}
\psline{-}(0,0)(0,-2)
\psline{-}(2,0)(2,-2)
\psarc{-}(4.5,0){0.5}{180}{360}
\psarc{-}(4.5,-2){0.5}{0}{180}
\end{pspicture} 
\right)
 i_N^d \hspace{-0.1cm}\left(
 \psset{unit=0.4}  
 \begin{pspicture}(-0.4,0)(5.4,0.8)
\rput(1.1,0){$\dots$}
\rput(3.1,0){$\dots$}
\rput(4,-0.6){$i$}
\rput(0,-0.6){$j$}
\rput(2,-0.6){$k$}
\psdots[{linewidth=0.7pt}](0,0)(2,0)(4,0)(5,0)
 \psset{linecolor=myc2}
\psbezier{-}(0,0)(0,2)(5,2)(5,0)
\psbezier{-}(2,0)(2,1.2)(4,1.2)(4,0)
\end{pspicture}\right)
  =  i_N^d \hspace{-0.1cm}\left(
 \psset{unit=0.4}  
 \begin{pspicture}(-0.4,0)(5.4,0.8)
\rput(1.1,0){$\dots$}
\rput(3.1,0){$\dots$}
\rput(4,-0.6){$i$}
\rput(0,-0.6){$j$}
\rput(2,-0.6){$k$}
\psdots[{linewidth=0.7pt}](0,0)(2,0)(4,0)(5,0)
 \psset{linecolor=myc2}
\psarc{-}(4.5,0){0.5}{0}{180}
\psbezier{-}(0,0)(0,1.2)(2,1.2)(2,0)
  \end{pspicture}\right)
    \rightarrow  e_i T_{j,i+1} T_{k,i} Y(v) |0\rangle = T_{i,i+1} T_{j,k} Y(v) |0\rangle. 
$ 
\end{itemize}
We now verify that each algebraic identity holds. Since $Y(v)$ commutes with $e_i$ and with $T_{i,j}$, $T_{i,k}$, ..., we can ignore it in our calculations. Because of (\ref{eq:eTL}), one can write
\begin{equation*}
e_j = \sigma^-_j\sigma^+_{j+1} + \sigma^+_j\sigma^-_{j+1} + (q+q^{-1}) \sigma^+_j \sigma^-_j\sigma^+_{j+1}\sigma^-_{j+1} - q \sigma^+_j \sigma^-_j - q^{-1}\sigma^+_{j+1}\sigma^-_{j+1}.
\end{equation*}
Since $\sigma^+_j \sigma^-_j | 0 \rangle = | 0 \rangle$ and $\sigma^+_j | 0 \rangle =  0$, it is obvious that 1) is satisfied. As opposed to the $\rho$ representation, here the number of defects is conserved, which explains the restriction $|_d$ given in the proposition. Similarly, for 2), 3) and 5),
\begin{align*} e_i T_{i,i+1} | 0 \rangle &= \big( w (\sigma_i^- \sigma_{i+1}^+ \sigma_{i+1}^- - q \sigma_i^+ \sigma_{i}^- \sigma_{i+1}^-) +w^{-1} (\sigma_i^+ \sigma_{i+1}^- \sigma_{i}^- - q^{-1} \sigma_{i+1}^+ \sigma_{i+1}^- \sigma_{i}^-) \big) |0 \rangle\\
& = -(q + q^{-1}) (w \sigma^-_{i+1} + w^{-1} \sigma^-_{i}) |0 \rangle= -(q + q^{-1}) T_{i,i+1} |0 \rangle, \\
 e_i T_{i+1,j} | 0 \rangle &= \big( w (0) +w^{-1} (\sigma^-_{i} \sigma^+_{i+1} \sigma^-_{i+1} -q \sigma^+_{i} \sigma^-_{i} \sigma^-_{i+1}) \big)|0 \rangle \\
& = (w^{-1} \sigma^-_i + w \sigma^-_{i+1}) |0 \rangle  = T_{i,i+1}|0 \rangle, \\
 e_i T_{i,k} T_{i+1,j}| 0 \rangle &= \big( w^2 (0) + w^{-2}(0)+w^{0} (\sigma_{i+1}^-\sigma_j^--q^{-1} \sigma_{i}^-\sigma_j^- + \sigma_i^-\sigma_k^- -q \sigma_{k}^-\sigma_{i+1}^-)\big)|0 \rangle \\
& = (w \sigma^-_{i+1} + w^{-1} \sigma^-_{i})(w \sigma^-_{k} + w^{-1} \sigma^-_{j}) |0 \rangle  = T_{i,i+1}T_{j,k}|0 \rangle.
\end{align*}
The proofs of 4), 6) and 7) do not require any new ideas and are left to the reader.
\hfill$\square$ 

The only difference between the action of the Temperley-Lieb algebra element $c$ on $V_N^d$ and that of the matrix $X(c)$ on $i_N^d(V_N^d)$ is that connected defects always give $0$ in the second case. Nevertheless, for any connectivity $c$, the diagonal blocks of $\rho(c)|_d$ can be calculated from those of $X(c)|_{S^z = d/2}$. Any information in non diagonal blocks in the loop model is lost in the XXZ model.

\subsection{The injectivity of $i^d_N$}
\begin{Definition} Path, Dyck path and order.
\begin{itemize}
\item[(a)] The set of paths with endpoint distance $y$, $P_y^N$, is the set of $\vec{x}=\{ x_1, x_2, ..., x_N\}$, where $x_i = \pm 1 \, \forall \, i$ and $\sum_{i=1}^N x_i =y $.
\item[(b)] The set of Dyck paths with endpoint distance $y$, $DP_y^N$, is the subset of $\vec{x}$ in $P_y^N$ satisfying  $\sum_{i=1}^k x_i \ge 0 $ for  all $k$ in $\{1,...,N \}$.
\item[(c)] We define an order for elements of $\vec{x} \in P_y^N$: $\vec{x}_1 < \vec{x}_2$ if $\mathcal{O}(\vec{x}_1) < \mathcal{O}(\vec{x}_2)$, with $\mathcal{O}(\vec{x}) = \sum_{i=1}^N 2^{i}\delta_{x_i,-1} $.
\end{itemize}
 
\label{sec:Dyckpath}
\end{Definition}

Dyck paths in $DP_y^N$ are paths starting from $(0,0)$ and ending at $(N,y)$ using steps $(1,1)$ and $(1,-1)$, that never venture in the lower half of the plane. The largest Dyck paths with respect to the ordering are those where the steps $(1,-1)$ are at the end of the path. One can easily be convinced that there are no $\vec{x}_1, \vec{x}_2$ in $DP_y^N$ such that  $\mathcal{O}(\vec x_1) = \mathcal{O}(\vec x_2)$ and $\vec x_1 \neq \vec x_2$.

Basis elements of $(\mathbb{C}^2)^{\otimes N}|_{S^z=N/2-n}$, labeled $|\alpha \rangle$, are vectors of length $N$ with every component $\in \{ +1,-1 \}$, indicating up and down spins. There exists a simple bijection between elements $\vec{x}$ in $P^{N-2n}$ and spin configurations $|\alpha \rangle$. To each path $\vec{x}$, we associate a configuration $\mathcal{C}(\vec{x})$: when $x_i$ = +1, the $i$-th spin is $\uparrow$, and when $x_i = -1$, $\downarrow$. 

\begin{Proposition} 
$i_N^d$ is injective.
\label{sec:injection}
\end{Proposition}
\noindent{\scshape Proof\ \ } Let the $v$s be elements of $B_N^{d=N/2-n}$ and the $| \alpha \rangle$s as before. To show that $i_N^d$ is injective, we must show that
$$ P_{\alpha,v} = \langle \alpha | i_N^d(v) \rangle,$$
a rectangular matrix of dimensions $\left( \begin{smallmatrix} N \\ n\end{smallmatrix}\right)$ by $\left(\begin{smallmatrix} N \\ n\end{smallmatrix}\right) - \left(\begin{smallmatrix} N \\ n-1\end{smallmatrix}\right)$  (and again $n=(N-d)/2$ is the number of bubbles), is of maximal rank. For this, we study a square matrix $\tilde{P}_{\alpha,v}$, of size $\left(\begin{smallmatrix} N \\ n\end{smallmatrix}\right) - \left(\begin{smallmatrix} N \\ n-1\end{smallmatrix}\right)$, with the same definition as $P_{\alpha,v}$, except we make a restriction on the spin configurations $| \alpha \rangle$. We will show that this matrix is of maximal rank. To this intent, we will order the $v$s of the link basis in decreasing order of their corresponding Dyck path, $\mathcal{O}(\mathcal{B}^{-1}(v))$ ($\mathcal{B}$ has been introduced in definition \ref{sec:deuxcolreduit}). For the $| \alpha \rangle$s, we choose the subset of spin configurations $| \alpha \rangle=\mathcal{C}(\vec{x})$ for $\vec{x}$ in $DP_y^N$, and order them, again, in decreasing order of $\mathcal{O}(\vec{x})$.

For a given $v \in B_N^d$, $\mathcal{C}(\mathcal{B}^{-1}(v))$ is the state in $(\mathbb C^2)^{\otimes N}|_{S^z=d/2}$ whose component in $i_N^d(v)$ has the biggest power of $w$: $n$. Indeed, $\mathcal{C}(\mathcal{B}^{-1}(v))$ is the configuration obtained by replacing every bubble of $v$ by $w \,  \uparrow \downarrow \, $. All other components of $i_N^d(v)$ are obtained from the first by replacing certain pairs $ \uparrow \downarrow \,$ by $ \downarrow \uparrow \, $ and by diminishing the power of $w$ by two for each pair changed. We conclude that in $\tilde{P}_{\alpha,v}$, every element on the diagonal is $w^n$ and is non zero (except for $w=0$, which is an unphysical case). Every component $| \alpha \rangle$ of $i_N^d(v)$ has a $\mathcal{O}(\mathcal{C}^{-1}(|\alpha\rangle))$ smaller or equal to $\mathcal{O}(\mathcal{B}^{-1}(v))$, and $\tilde{P}_{\alpha,v}$ matrix is therefore lower triangular. From the previous remark, the rank of $\tilde{P}_{\alpha,v}$, and therefore of $P_{\alpha,v}$, is maximal.

\hfill$\square$ \\
An example, with $N=5$, $n=2$, $d=1$: \\
\begin{tabular}{ c c c c c c c}
  $\vec{x} \in DP^5_1$: & \qquad &  
\psset{unit=0.8}
\begin{pspicture}(0,-0)(1,1.0)
\psset{linewidth=1pt}
\psline{-}(0,0)(0.2,0.2)(0.4,0.4)(0.6,0.6)(0.8,0.4)(1,0.2)
\psdots[dotsize=0.10](0,0)(0.2,0.2)(0.4,0.4)(0.6,0.6)(0.8,0.4)(1,0.2)
\end{pspicture}
  & 
\psset{unit=0.8}  
\begin{pspicture}(0,-0)(1,1)
\psset{linewidth=1pt}
\psline{-}(0,0)(0.2,0.2)(0.4,0.4)(0.6,0.2)(0.8,0.4)(1,0.2)
\psdots[dotsize=0.10](0,0)(0.2,0.2)(0.4,0.4)(0.6,0.2)(0.8,0.4)(1,0.2)
\end{pspicture}
  &
  \psset{unit=0.8}  
\begin{pspicture}(0,-0)(1,1)
\psset{linewidth=1pt}
\psline{-}(0,0)(0.2,0.2)(0.4,0)(0.6,0.2)(0.8,0.4)(1,0.2)
\psdots[dotsize=0.10](0,0)(0.2,0.2)(0.4,0)(0.6,0.2)(0.8,0.4)(1,0.2)
\end{pspicture}
  &
  \psset{unit=0.8}  
\begin{pspicture}(0,-0)(1,1)
\psset{linewidth=1pt}
\psline{-}(0,0)(0.2,0.2)(0.4,0.4)(0.6,0.2)(0.8,0.0)(1,0.2)
\psdots[dotsize=0.10](0,0)(0.2,0.2)(0.4,0.4)(0.6,0.2)(0.8,0.0)(1,0.2)
\end{pspicture}
  &
  \psset{unit=0.8}  
\begin{pspicture}(0,-0)(1,1)
\psset{linewidth=1pt}
\psline{-}(0,0)(0.2,0.2)(0.4,0.0)(0.6,0.2)(0.8,0.0)(1,0.2)
\psdots[dotsize=0.10](0,0)(0.2,0.2)(0.4,0.0)(0.6,0.2)(0.8,0.0)(1,0.2)
\end{pspicture} 
  \\
  &&&& & &
  \\
  $\mathcal{O}(\vec{x})$ &&$2^4+2^5$&$2^3+2^5$&$2^2+2^5$& $2^3+2^4$& $2^2+2^4$
  \\
  &&&& & &
  
   \\
  $\mathcal{B}(\vec{x}) \in V_5^1$: & \qquad & 
\begin{pspicture}(0,-0)(1,0.4)
\psset{linewidth=1pt}
\psdots[dotsize=0.08](0.1,0)(0.3,0.0)(0.5,0.0)(0.7,0.0)(0.9,0.0)
\psset{linecolor=myc2}
\psline{-}(0.1,0)(0.1,0.4)
\psarc{-}(0.6,0){0.1}{0}{180}
\psbezier{-}(0.3,0)(0.3,0.3)(0.9,0.3)(0.9,0)
\end{pspicture}
 & 
\begin{pspicture}(0,-0)(1,0.4)
\psset{linewidth=1pt}
\psdots[dotsize=0.08](0.1,0)(0.3,0.0)(0.5,0.0)(0.7,0.0)(0.9,0.0)
\psset{linecolor=myc2}
\psline{-}(0.1,0)(0.1,0.4)
\psarc{-}(0.4,0){0.1}{0}{180}
\psarc{-}(0.8,0){0.1}{0}{180}
\end{pspicture}
& 
\begin{pspicture}(0,-0)(1,0.4)
\psset{linewidth=1pt}
\psdots[dotsize=0.08](0.1,0)(0.3,0.0)(0.5,0.0)(0.7,0.0)(0.9,0.0)
\psset{linecolor=myc2}
\psline{-}(0.5,0)(0.5,0.4)
\psarc{-}(0.8,0){0.1}{0}{180}
\psarc{-}(0.2,0){0.1}{0}{180}
\end{pspicture}
& 
\begin{pspicture}(0,-0)(1,0.4)
\psset{linewidth=1pt}
\psdots[dotsize=0.08](0.1,0)(0.3,0.0)(0.5,0.0)(0.7,0.0)(0.9,0.0)
\psset{linecolor=myc2}
\psline{-}(0.9,0)(0.9,0.4)
\psarc{-}(0.4,0){0.1}{0}{180}
\psbezier{-}(0.1,0)(0.1,0.3)(0.7,0.3)(0.7,0)
\end{pspicture}
&
\begin{pspicture}(0,-0)(1,0.4)
\psset{linewidth=1pt}
\psdots[dotsize=0.08](0.1,0)(0.3,0.0)(0.5,0.0)(0.7,0.0)(0.9,0.0)
\psset{linecolor=myc2}
\psline{-}(0.9,0)(0.9,0.4)
\psarc{-}(0.2,0){0.1}{0}{180}
\psarc{-}(0.6,0){0.1}{0}{180}
\end{pspicture}\\
  &&&&&&
  \\
  $\mathcal{C}(\vec{x}) \in (\mathbb{C}^2)^{\otimes 5}$: & \qquad & $| \uparrow \uparrow \uparrow \downarrow \downarrow \, \rangle$ & $| \uparrow \uparrow  \downarrow \uparrow \downarrow \, \rangle$ & $| \uparrow \downarrow \uparrow \uparrow \downarrow \, \rangle$ & $| \uparrow \uparrow \downarrow \downarrow \uparrow \, \rangle$ & $| \uparrow \downarrow \uparrow \downarrow \uparrow \, \rangle$ \\
\end{tabular}
\begin{equation*}
\tilde{P}_{\alpha,v} = \begin{pmatrix}
w^2 & 0 & 0& 0& 0\\
1& w^2& 0& 0& 0\\
0& 1& w^2& 0& 0\\
0& 1& 0& w^2& 0\\
1& w^{-2}& 1& 1& w^2\\
\end{pmatrix}.
\end{equation*}

From propositions \ref{sec:fermeture} and \ref{sec:injection},  $i_N^d(V_N^d)$ is a subspace of $\textrm{dim}V_N^d$ of $(\mathbb{C}^2)^{\otimes N}|_{S^z = d/2}$, invariant under the action of the $e_i$s of XXZ. The eigenvectors of $\rho(\mathcal{H}_N)$ (for any $\beta$), restricted to the sector with $d$ defects, are in correspondence with eigenvectors of $H_{XXZ}$ in the $S^z = d/2$ sector.

\subsection{The relation between $U_q(sl_2)$ and $i_N^d(V_N^d)$}
In this section, we establish the relation between the homomorphism $i_N^d$ and the algebra $U_q(sl_2)$.

\begin{Proposition}
For all $v \in V_N^d$, $ i_N^d(v) \in \ker \,S^+$. 
\label{sec:kersplus}
\end{Proposition}
\noindent{\scshape Proof\ \ }
We start by restricting the proof to link patterns with only simple bubbles, i.e. \hspace{-0.1 cm}to $v$s for which every $(i,j) \in \psi(v)$ is of the form $(i,i+1)$. We notice that, in general, whenever $k$ does not appear in any of the pairs $(i,j)$ in $\psi(v)$, $S^+_k  i_N^d(v) = 0$. Indeed, when $i \neq j$,
$$ S^+_i \sigma_j^- = q^{s_{i,j}}  \sigma_j^- S^+_i \qquad \textrm{where} \qquad
s_{i,j} = \left\{ 
\begin{array}{l l}
  +1, & \quad \textrm{if} \quad i>j, \\
  -1, & \quad \textrm{if}\quad i<j. \end{array} \right.  $$
and, when $k$ is not in any of the pairs of  $\psi(v)$,
\begin{align*}
S^+_k  i_N^d(v) &= S^+_k \left( \prod_{(i,i+1) \in \psi(v)}  (w \sigma^-_{i+1} + w^{-1} \sigma^-_i) \right) | 0 \rangle \\
&=  \left( \prod_{(i,i+1) \in \psi(v)}  (q^{s_{k,i+1}}w \sigma^-_{i+1} + q^{s_{k,i}}w^{-1} \sigma^-_i) \right) S^+_k | 0 \rangle = 0. 
\end{align*}
All that is left to calculate is $ S^+ i_N^d(v) = \sum_{(k,k+1) \in \psi(v)} (S^+_{k} + S^+_{k+1})  i_N^d(v)$, 
\begin{align*}
(S^+_{k} + S^+_{k+1})  i_N^d(v) &=  \left( \displaystyle\prod_{\substack{  (i,i+1) \in \psi(v) \\  i\neq k}}  (q^{s_{k,i+1}}w \sigma^-_{i+1} + q^{s_{k,i}}w^{-1} \sigma^-_i) \right) S^+_{k} (w \sigma^-_{k+1} + w^{-1} \sigma^-_{k})| 0 \rangle \\
& \qquad + \left( \displaystyle\prod_{\substack{  (i,i+1) \in \psi(v) \\  i\neq k}}  (q^{s_{k+1,i+1}}w \sigma^-_{i+1} + q^{s_{k+1,i}}w^{-1} \sigma^-_i) \right) S^+_{k+1} (w \sigma^-_{k+1} + w^{-1} \sigma^-_{k})| 0 \rangle \\
&=  \left( \displaystyle\prod_{\substack{  (i,i+1) \in \psi(v) \\  i\neq k}} q^{s_{k,i}} (w \sigma^-_{i+1} + w^{-1} \sigma^-_i) \right) (S^+_{k}+S^+_{k+1}) (w \sigma^-_{k+1} + w^{-1} \sigma^-_{k})| 0 \rangle.
\end{align*}
When $v$ has only simple bubbles, $s_{k,i} = s_{k+1,i} = s_{k,i+1} = s_{k+1,i+1}$. This has been used at the last equality. Finally,
\begin{align*}
 (S^+_{k}&+S^+_{k+1})  (w \sigma^-_{k+1} + w^{-1} \sigma^-_{k})| 0 \rangle =  w^{-1} S^+_{k}  \sigma^-_{k}| 0 \rangle + w S^+_{k+1} \sigma^-_{k+1} | 0 \rangle \\
 &= w^{-1} \left( \prod_{i=1}^{k-1} q^{-\sigma^z_i/2} \right) \sigma^+_{k}\sigma^-_{k} \left( \prod_{j=k+1}^{N} q^{\sigma^z_j/2} \right) | 0 \rangle + w \left( \prod_{i=1}^{k} q^{-\sigma^z_i/2} \right) \sigma^+_{k+1}\sigma^-_{k+1} \left( \prod_{j=k+2}^{N} q^{\sigma^z_j/2} \right) | 0 \rangle \\
 & = w^{-1} \left( q^{(N-2k+1)/2} + w^2 q^{(N-2k-1)/2} \right) | 0 \rangle  = 0. \end{align*}

For $w \in B_N^d$ with bubbles that are not simple, from proposition \ref{sec:fermeture}, one can write $i_N^d(w) = (\prod_{ j \in J} e_{j}) i^d_n(v)$ for some set $J$ and for $v$ a link state with only simple bubbles. Since $[S^+,e_j] = 0$ by proposition \ref{sec:Uqcommutes}, $S^+ i_N^d(w) = 0$ for all $w \in B_N^d$.

\hfill$\square$ \\
From this proposition, it follows that for $q=q_c$ and $(q_c)^{2P}=1$, $ i_N^d(V_N^d)$ is also $\subset \ker \,S^{+(P)}$:

\begin{equation*}
S^{+(P)}  i_N^d(v) = \lim_{q \rightarrow q_c} \frac{(S^{+})^P  i_N^d(v)}{[P]_q} =  \lim_{q \rightarrow q_c} \frac{0}{[P]_q} = 0.
\end{equation*}
\section{The reduction of state space and the degeneracies}
%
\label{sec:reduction}
In the last sections, we found that the set of eigenvalues of $\rho(\mathcal{H}_N)$ in the sector with $n$ bubbles was a subset of the eigenvalues of $H$ in the sector $S^z = N/2-n$. For $\beta = 0$, this will allow us to prove the selection rules of section \ref{sec:intro2}: we will calculate the degeneracy of every eigenvalue in $H$, remove those that are tied to eigenvectors not in $i_N^d(V_N^d)$ and show that the degeneracy obtained match those of the loop model, given by eqs (\ref{eq:conjequiv}), (\ref{eq:conjequiv2}) and (\ref{eq:conjequiv3}). The two corresponding vector spaces  $(\mathbb{C}^2)^{\otimes N}|_{S^z = N/2-n}$ and $V_N^{N-2n}$ have respective dimensions $\left(\begin{smallmatrix} N \\ n\end{smallmatrix}\right)$ and $\left(\begin{smallmatrix} N \\ n\end{smallmatrix}\right) - \left(\begin{smallmatrix} N \\ n-1\end{smallmatrix}\right)$. To get only states in $i_N^d(V_N^{N-2n})$, we will need to remove $\left(\begin{smallmatrix} N \\ n-1\end{smallmatrix}\right)$ independent states from $(\mathbb{C}^2)^{\otimes N}|_{S^z = N/2-n}$.

\begin{Definition} Let $\mathcal{O} = \sum_{\vec{i}} \alpha_{\vec{i}} \mathcal{O}_{\vec{i}}$ with $\vec{i} = (i_1, i_2, ... i_{|\vec{i}|})$, where $\alpha_{\vec{i}} \in \mathbb{C}$ and  $\mathcal{O}_{\vec{i}}$ is the product of some annihilation operators:  $\mathcal{O}_{\vec{i}} = \prod_{k=1}^{|\vec{i}|} b_{i_k}$. We define $\mathcal{O}'$ with the following two rules:
\begin{itemize}
\item $\mathcal{O}' =  \sum_{\vec{i}} \alpha^*_{\vec{i}} \mathcal{O}'_{\vec{i}}$,
\item  $\mathcal{O}'_{\vec{i}} =  \prod_{k=1}^{|\vec{i}|} a'_{ i_{|\vec{i}| +1 -k}}$,  
\end{itemize}
where the product of non-commuting elements is always taken from left to right.
\label{sec:Oprime}
\end{Definition}

\noindent The sum over $ \vec{i} $ is a sum over multi-indexes that could potentially have different lengths, but the only $\mathcal{O}$s we will need have $\mathcal{O}_{\vec{i}}$ with a unique fixed length. Examples:
$$ (b_3 b_6 b_1)' = a_1 a_6 a_3,  \qquad (3i b_2 + (5i+1) b_7 b_4 + 12 b_0 b_2 b_1 b_{12})' =  -3i a_2 + (-5i+1) a_4 a_7 + 12 a_{12}a_1 a_2 a_0.$$


\begin{Proposition} Let an operator $\mathcal O \neq 0$ that satisfies $\mathcal O i_N^d (v) = 0$ for all $v \in V_N^d$. Then $\mathcal O' |0\rangle \notin i_N^d (V_N^d)$.
\end{Proposition}
\noindent{\scshape Proof\ \ }
There does not exist a set of constants $\gamma_v$s such that
$$ \mathcal O' |0 \rangle + \sum_{v \in V_N^d} \gamma_v i_N^d(v) = 0.$$
Indeed, multiplying this equation from the left with $\mathcal O$, the second term is zero by hypothesis, 
$$ \mathcal O \mathcal O' |0 \rangle  = \sum_{\vec{i}} |\alpha_{\vec{i}}|^2 |0 \rangle = 0, $$
which contradicts the hypothesis $\mathcal O \neq 0$.
\hfill$\square$ \\

By proposition \ref{sec:kersplus}, the operators $S^+$ and $S^{+(2)}$ are two such operators $\mathcal O$ satisfying $\mathcal{O} i_N^d(v) = 0,  \forall \, v \in V_N^d$. To find eigenvectors of $H$ not in $i_N^d(V_N^d)$ and that we will have to remove from all the states of the form $a_{i_1}a_{i_2}...a_{i_n} |0 \rangle$ (with $n = (N-d)/2$), we look for operators $\mathcal{O} = \sum_{\vec i} \alpha_{\vec i} \mathcal O_{\vec i}$ for which every $\mathcal{O}_{\vec i}$ is a product of $n$ annihilation operators. They are:
\begin{itemize}
\item $ \mathcal{O} = S^+ b_{j_1}b_{j_2}...b_{j_{n-1}} $ where $j_k \neq 0$ for $k = 1, ..., n-1$ ($b_0$ is the generator corresponding to $S^+$, see eqs (\ref{eq:Sodd}) and (\ref{eq:Seven}), and $\mathcal{O}$ must be non zero). Because $\{b_0,b_j \} = 0$ for all $j$, $\mathcal{O} i_N^d(v) = 0$ for all $v$. All the states  $a_{j_{n-1}}a_{j_{n-2}}...a_{j_{1}} a_0 |0 \rangle$ must be removed. They will be referred to as states {\it of the first kind}. There are $\left(\begin{smallmatrix} N-1 \\ n-1 \end{smallmatrix}\right)$ such states.
\item  $ \mathcal{O} = S^{+(2)} b_{k_1}b_{k_2}...b_{k_{n-2}}$, and $\mathcal{O} i_N^d(v) = 0$ for all $v$ by the same argument. The states to be removed are of the form $a_{k_{n-2}}a_{k_{n-3}}...a_{k_{1}} (S^{+(2)})' |0 \rangle$, where the $a_{k_i}$s can be any of the $N-1$ remaining operators (not $a_0$, as we want to avoid any overlap with states of the first kind). They will be referred to as states {\it of the second kind}. There are $\left(\begin{smallmatrix} N-1 \\ n-2 \end{smallmatrix}\right)$ such states.
\end{itemize}
Of course, $\left(\begin{smallmatrix} N-1 \\ n-1 \end{smallmatrix}\right) + \left(\begin{smallmatrix} N-1 \\ n-2 \end{smallmatrix}\right) = \left(\begin{smallmatrix} N \\ n-1 \end{smallmatrix}\right)$, precisely the number of states we need to remove. That all these states are independent is non trivial and shown in appendix \ref{app:b}. Having succeeded in finding a rule that removes all eigenstates of $H$ not in $i_N^d(V_N^d)$, we can now calculate the degeneracies.

\subsection{$N$ odd}
As seen in section \ref{sec:diagNimpair}, when $N$ is odd, the eigenvectors of $H$, restricted to the $S^z = N/2 - n$ sector, are of the form
\begin{equation} |\gamma \rangle = \left( \prod_{i=1}^n a_{k_i} \right)  |0\rangle \label{eq:allvects}\end{equation}
for $k_i \in \{0, 1, ..., N-1\}$. If one of the $k_i$s is $0$, we put is at the end and set $a_{k_n} = a_0$. The eigenvalues are
\begin{itemize}
\item[(a)] $\gamma =  2\sum_{i=1}^{n} \cos (\pi k_i/N)$, if no $k_i$ is $0$,
\item[(b)] $\gamma =  2\sum_{i=1}^{n-1} \cos (\pi k_i/N)$, if some $k_i$ is $0$.
\end{itemize}
We call $\Gamma_0^n$ and $\Gamma_1^n$ respectively the set of all $\gamma$s  for (a) and (b).

\begin{Proposition} $ \Lambda_\delta^n = \Gamma_\delta^n $ for both $\delta = 0$ and $1$.
\label{sec:inclusion1}\end{Proposition}
\noindent{\scshape Proof\ \ } Let $\gamma \in \Gamma_\delta^n$. To show that $\gamma \in \Lambda_\delta^n$, we construct the three subsets $K^+$, $K^-$ and $K^c$. For all $k \in \{1, ..., (N-1)/2\}$,
\begin{itemize}
\item[(i)] if $k \in \{ k_1, ..., k_n\}$ and $N-k \notin \{ k_1, ..., k_n\}$, we put $k$ in $K^+$;
\item[(ii)] if $k \notin \{ k_1, ..., k_n\}$ and $N-k \in \{ k_1, ..., k_n\}$, we put $k$ in $K^-$;
\item[(iii)] if $k \in \{ k_1, ..., k_n\}$ and $N-k \in \{ k_1, ..., k_n\}$, we put $k$ in $K^c$;
\item[(iv)] if $k \notin \{ k_1, ..., k_n\}$ and $N-k \notin \{ k_1, ..., k_n\}$, we put $k$ in $K^c$;
\end{itemize}
We stress that when $k_n = 0$, $0$ is not in any of $K^+$, $K^-$ or $K^c$, but for fixed $n$, its presence or absence changes the number of elements in $K^+ \cup K^-$. The case  $\delta = 0$ is when the $a_0$ creation operator is absent: $n-m = n-|K^+ \cup K^- |$ counts the number of elements in  (iii) and is even. When $\delta=1$, the last momentum is $k_n=0$ and the number of elements in (iii) is still even, but now given by $n-1-m$, so $n-m$ is odd. 
\\

\noindent Now, let $\lambda \in \Lambda^n_\delta$ with a fixed $m$. To see it is also in $ \Gamma_\delta^n$, we construct the set of momenta as follows
\begin{itemize}
\item[(i)] if $k$ is in $K^+$, we put $k$ in  $\{ k_1, ..., k_n\}$, but not $N-k$;
\item[(ii)] if $k$ is in $K^-$, we put $N-k$ in  $\{ k_1, ..., k_n\}$, but not $k$;
\item[(iii)] if $\delta=1$, we set $k_n =0$, 
\item[(iv)] For all the $k$s that are in $K^c$, we choose $(n-m - \delta)/2$ among the $(N-1)/2-m$ remaining values and put, for each, $k$ and $N-k$ in $\{ k_1, ..., k_n\}$.
\end{itemize}
\hfill$\square$

\noindent From the previous construction, an eigenvalue $\lambda$ of $H$ has eigenvector 
\begin{align} &\left( \prod_{i} a_{N-i} a_{i} \right) \left( \prod_{j \in K^-} a_{N-j} \right) \left( \prod_{k \in K^+} a_{k} \right)  |0\rangle, \quad \textrm{if} \quad \delta = 0, \label{eq:allvect1}\\
&\left( \prod_{i} a_{N-i} a_{i} \right) \left( \prod_{j \in K^-} a_{N-j} \right) \left( \prod_{k \in K^+} a_{k} \right)  a_0 |0\rangle, \quad \textrm{if} \quad \delta = 1. \label{eq:allvect2} \end{align}
 where the product on $i$ has $(n-m-\delta)/2$ terms, all different, with $i \in  K^c$. The degeneracy comes from all the possibilities for the product on $i$, and is given by $$\textrm{deg}_{H}(\lambda) = \begin{pmatrix} \frac{N-1}{2}-m \\ \frac{n-m-\delta}{2}\end{pmatrix}.$$
 To obtain the degeneracies of these eigenvalues in $\rho(\mathcal{H}_N)$, we remove the states of (\ref{eq:allvect2}) (they are all of the first kind) and from (\ref{eq:allvect1}), all the states of the second kind,
  \begin{equation} \left( \prod_{i'} a_{N-i'} a_{i'} \right) \left( \prod_{j \in K^-} a_{N-j} \right) \left( \prod_{k \in K^+} a_{k} \right) (\sum_{l=1}^{(N-1)/2} a_l a_{N-l}) |0\rangle, \label{eq:vects2}\end{equation}
 where the products on $i'$ has $(n-m-2)/2$ terms and where the constant $(-1)^{S^z}$ of (\ref{eq:S+-odd}) has been dropped for convenience. For some $\lambda$ with a fixed value of $m$, there are $\left(\begin{smallmatrix} \frac{N-1}{2}-m \\ \frac{n-m-2}{2}\end{smallmatrix}\right)$ such possible choices, each corresponding to an eigenvector. The set of corresponding eigenvectors is linearly independent (see appendix \ref{app:b}) and the result is
  \begin{itemize}
  \item For $\lambda \in  \Lambda_0^n$, $\textrm{deg}_{\mathcal{H}}(\lambda) = \left( \begin{smallmatrix} \frac{N-1}{2}-m \\ \frac{n-m}{2}\end{smallmatrix}\right) - \left(\begin{smallmatrix} \frac{N-1}{2}-m \\ \frac{n-m-2}{2}\end{smallmatrix}\right)$,
  \item For $\lambda \in  \Lambda_1^n$, $\textrm{deg}_{\mathcal{H}}(\lambda) = 0$.
  \end{itemize}
  This is precisely the content of conjecture \ref{sec:degimpair}, which is now proved.
  
\subsection{$N$ even}
As in section \ref{sec:diagNimpair}, eigenvectors and generalized eigenvectors of $H$, for $S^z = N/2 - n$, are given in (\ref{eq:allvects}), but with $k_i \in \{-1, 0,1, ..., N-1\} \setminus \{N/2 \}$. When the $a_0$ and/or $a_{-1}$ excitations are present, we set them to the last $k_i$s ($k_n$ and $k_{n-1}$, when both are present). Eigenvalues are
\begin{itemize}
\item[(a)] $\gamma =  2\sum_{i=1}^{n} \cos (\pi k_i/N)$ if $a_0, a_{-1}$ are not in the $a_i$s;
\item[(b)] $\gamma =  2\sum_{i=1}^{n-1} \cos (\pi k_i/N)$ if 
	\begin{itemize}
	\item[(i)] $a_{-1}$ is not in the $a_i$s, but $a_0$ is;
	\item[(ii)] $a_0$ is not in the $a_i$s, but $a_{-1}$ is;
	\end{itemize}
\item[(c)] $\gamma = 2\sum_{i=1}^{n-2} \cos (\pi k_i/N)$ if $a_0$ and $a_{-1}$ are both among the $a_i$s.
\end{itemize}
We refer to the sets of eigenvalues in the cases (a), (b) and (c) as $\Gamma_a^n$, $ \Gamma_b^n$ and $ \Gamma_c^n$.
 
\begin{Proposition} Based on the definition of \ref{sec:enslambimpair} for $\Lambda_0^n $ and $\Lambda_1^n$, $\Gamma_a^n = \Lambda^n_0$, $\Gamma_b^n = \Lambda^n_1$ and $\Gamma_c^n \subset \Lambda^n_0$.
 \label{sec:inclusion2}\end{Proposition}
The proof is identical to the proof of proposition \ref{sec:inclusion1}, with a few subtleties. The first is that whenever $\gamma$ has the $a_{-1}$ excitation, the $a_{0}$ excitation or both, their momenta are not in either $K^+$, $K^-$ or $K^c$, but their absence changes the number of elements in $K^+\cup K^-$. The second concerns the fact that $\Gamma_c^n$ is only a subset of $\Lambda^n_0$. Indeed, the elements of $\Lambda^n_0$ with $m=n$ are not contained in $\Gamma^n_c$. The rest of the proof is not repeated. Note that the number of pairs $(k,N-k)$ to be fixed (among the $(N-2)/2-m$ possible choices) and the degeneracies of the eigenvalues are different for the three cases (a), (b) and (c):
\begin{itemize} 
\item[(a)] $(n-m)/2$ pairs to be fixed and $\textrm{deg}_{H}(\lambda)=\left(\begin{smallmatrix} \frac{N-2}{2}-m \\ \frac{n-m}{2}\end{smallmatrix}\right)$;
\item[(b)] $(n-m-1)/2$ pairs to be fixed and $\textrm{deg}_{H}(\lambda)=\left(\begin{smallmatrix} \frac{N-2}{2}-m \\ \frac{n-m-1}{2}\end{smallmatrix}\right)$;
\item[(c)] $(n-m-2)/2$ pairs to be fixed and $\textrm{deg}_{H}(\lambda)=\left(\begin{smallmatrix} \frac{N-2}{2}-m \\ \frac{n-m-2}{2}\end{smallmatrix}\right)$.
\end{itemize}
States to be removed are those of the first kind, see (\ref{eq:allvect2}), and those of the second kind,
  \begin{equation*} \left( \prod_{i'} a_{N-i'} a_{i'} \right) \left( \prod_{j \in K^-} a_{N-j} \right) \left( \prod_{k \in K^+} a_{k} \right) (\sum_{l=1}^{(N-2)/2} a_l a_{N-l}) |0\rangle, \end{equation*}
 and the product on $i'$ has $(n-m-2)/2$ terms. The $a_0a_{-1}$ contribution from $(S^{+(2)})'$ has been removed because this caused an overlap with states of the first kind. The degeneracies are
\begin{itemize}
\item[(a)] $\textrm{deg}_{\mathcal{H}} = \left(\begin{smallmatrix} \frac{N-2}{2}-m \\ \frac{n-m}{2}\end{smallmatrix}\right) - \left(\begin{smallmatrix} \frac{N-2}{2}-m \\ \frac{n-m-2}{2}\end{smallmatrix}\right)$,
\item[(b)] 
	\begin{itemize}
	\item[(i)]  $\textrm{deg}_{\mathcal{H}} = 0$,
	\item[(ii)]  $\textrm{deg}_{\mathcal{H}} = \left(\begin{smallmatrix} \frac{N-2}{2}-m \\ \frac{n-m-1}{2}\end{smallmatrix}\right) - \left(\begin{smallmatrix} \frac{N-2}{2}-m \\ \frac{n-m-3}{2}\end{smallmatrix}\right)$,
	\end{itemize}
\item[(c)]  $\textrm{deg}_{\mathcal{H}} =0$.
\end{itemize}
The cases (a) and (c) correspond to $\Lambda^n_0$, while (b)(i) and (b) (ii) correspond to $\Lambda^n_1$. This is the result of conjecture \ref{sec:degpair} and concludes the proof of the selection rules.

Note that Jordan partners were the states of (b)(i). Since they have all been removed, $\rho(\mathcal{H}_N)$ is diagonalizable.
%
\section{Conclusion} \label{sec:conclusion}
%

In this paper, we proved that the degeneracies of the eigenvalues of $\rho(\mathcal H_N)$, as given by the selection rules, are correct. We must stress however that the proof ignored the problem of accidental degeneracies resulting from accidental trigonometric identities. Another problem is the case of loop models on other geometries. A recent paper \cite{PRV} solved the model of critical dense polymers on the cylinder. An inversion relation was computed, eigenvalues were found and degeneracies conjectured by different selection rules from the ones here. The method proposed here might also lead to a proof of these conjectures.

%
%

\section*{Acknowledgements} 
The author thanks the NSERC for an Alexander Graham Bell graduate fellowship. The author would like to thank Yvan Saint-Aubin for fruitful discussions throughout the project and for reviewing his manuscript. He has benefitted from discussions with J\o rgen Rasmussen and David Ridout, and would like to thank Michael Doob for sharing his knowledge of graph theory.

%
%

\bigskip

\bigskip

\noindent{\LARGE\bfseries Appendices}

%
%

\appendix\section{The computation of $K'_{N/2}$, $\beta_1$ and $\beta_2$ (for $N$ even)}\label{app:a}

The goal of this section is to calculate the three constants $K'_{N/2}$, $\beta_1$ and $\beta_2$ that fix the two states (the eigenstate and its Jordan partner) tied to the eigenvalues $\xi = 0$ of $\mathcal N$. The anticommutation relations, in terms of $u_{N/2}$ and $w$, are rewritten as
$$ \{b_{-1},a_{-1}\} = \vec{f}_{-1}^T \vec{g}_{-1} = \sum_{j=1}^N u_{N/2}^j w^j = 1,$$
$$ \{b_{0},a_{-1}\} = \vec{f}_{0}^T \vec{g}_{-1} = \sum_{j=1}^N (u_{N/2}^j)^2 =0,$$
$$ \{b_{-1},a_{0}\} = \vec{f}_{-1}^T \vec{g}_{0} = \sum_{j=1}^N (w^j)^2 =0,$$
$$ \{b_{0},a_{0}\} = \vec{f}_{0}^T \vec{g}_{0} =\sum_{j=1}^N w^j u_{N/2}^j =1.$$
The second relation is trivially satisfied, since $\sum_{j=1}^N (-1)^j =0$ for $N$ even. The third constraint reads
\begin{equation} \beta_1^2 w_1^T w_1 + \beta_2^2 w_2^T w_2 + 2 \beta_1\beta_2 w_1^Tw_2 =0. \label{eq:thirdconst}\end{equation}
To calculate $w_1^T w_1$,
\begin{align*}
w_1^T w_1 &= \sum_{j=1}^N (-1)^j \Big\lfloor \frac{N-j-1}{2}\Big\rfloor^2 \\
		&= \sum_{k=1}^{N/2} \left((-1)^{2k} \Big\lfloor \frac{N-2k-1}2 \Big\rfloor^2 + (-1)^{2k-1} \Big\lfloor \frac{N-2k}2 \Big\rfloor^2\right)\\
		&= \sum_{k=1}^{N/2}  (N/2-k-1)^2 - \sum_{k=1}^{N/2}  (N/2-k)^2
		= -N(N-4)/4,
\end{align*}
and one can also find $w_2^T w_2 = -N^2/4$, $w_1^T w_2 = -N(N-2)/4$, and, from (\ref{eq:thirdconst}), $\beta_2/\beta_1 = -(N-4)/N$ (the second solution, $\beta_2/\beta_1 = -1$, is not retained, because it corresponds to the eigenvector $u_{N/2}^j = K'_{N/2}(w_2^j-w_1^j) = K'_{N/2}i^{j}$ that we already found). It only remains to fulfill the first constraint (the fourth one is identical):
\begin{align*}
 1 & = \sum_{j=1}^N u_{N/2}^j w^j = K'_{N/2} \beta_1 (-w_1^Tw_1 - \frac{N-4}N  w_2^Tw_2 + (\frac{N-4}N + 1)w_2^Tw_1)\\
    & = K'_{N/2} \beta_1 \left( \frac{N(N-4)}{4} + \frac{N(N-4)}{4} - (\frac{N-4}N + 1)\frac{N(N-2)}4  \right)
     = -2 K'_{N/2} \beta_1
\end{align*}
which gives $K'_{N/2} \beta_1 = -1/2$. Finally, a last constraint is obtained from $\mathcal{N} \vec{g}_{0} = \vec{g}_{-1}$, which is equivalent to imposing that the coefficient in front of $b_{0}a_{-1}$ is $1$ in eq. (\ref{eq:Hnondiag}):
$$ K'_{N/2} i^j = g_{-1}^j = (\mathcal{N}\vec{g}_{0})^j = \beta_1 (\mathcal{N} ( w_1 - w_2 (N-4)/N ))^j = \beta_1 i^{j-1} (1-(N-4)/N) = \beta_1 i^j (-4i/N) $$
where eq. (\ref{eq:actionNw}) has been used at the fourth equality. This gives $K'_{N/2}/\beta_1 = -4i/N$ and the calculation of the three constants is complete.
\medskip

%
%

\section{Independence of states not in $ i_N^d(V_N^d)$}\label{app:b}

In section \ref{sec:reduction}, we have identified states to be removed from $(\mathbb C^2)^{\otimes N}|_{S^z = d/2}$ and that should form a basis for the complement of $i_N^d(V_N^d)$. In this section, we show these states are non zero and independant. 

\begin{Definition}
Let $|v_1\rangle$ and $|v_2\rangle$ be any vector that can be written as $\mathcal O_1 |0\rangle$ and $\mathcal O_2 |0\rangle$, where $\mathcal O_1$ and $\mathcal O_2$ are multi-indexes as in definition \ref{sec:Oprime}. We introduce a scalar product between such states by defining $(|v_1\rangle, |v_2\rangle) = \langle0| \mathcal O_1' \mathcal O_2 |0\rangle$. We will denote this scalar product by $\langle v_1|v_2\rangle$.
\end{Definition}

The fact that states of the first kind $|w\rangle=a_{j_{n-1}}a_{j_{n-2}}...a_{j_{1}} a_0 |0 \rangle$ (with $j_1<j_2< ... <j_{n-1}$) are independent and non zero is trivial, as the scalar product restricted to such states is just $\langle w_1 | w_2 \rangle = \delta_{w_1,w_2}$: they all have length one and are mutually orthogonal. There are $\left( \begin{smallmatrix} N-1 \\ n-1 \end{smallmatrix} \right)$ such vectors.

The proof for vectors of the second kind is more involved. It requires the following definition (\cite{BCN},\cite{Godsil2}).
\begin{Definition}
Let $v$ and $k$ be positive integers, with $v>k$. The Johnson graph $J(v,k)$ is the following:
\begin{itemize}
\item its vertices $\theta$ are the subsets of length $k$ of $\{ 1, 2, ..., v \}$, their number is $\left( \begin{smallmatrix} v \\ k \end{smallmatrix} \right)$;
\item two vertices $\theta_1$ and $\theta_2$ are connected by an edge if and only if $|\theta_1 \cap \theta_2| = k-1$.
\end{itemize}
The adjacency matrix $A(v,k)$ of the Johnson graph $J(v,k)$ is the matrix with entries
\begin{equation*} A(v,k)_{\theta_1,\theta_2} = \left\{ \begin{array}{l l} 1 & \quad \textrm{if $\theta_1$ and $\theta_2$ are connected by an edge} ,\\ 0& \quad  \textrm{otherwise (even if $\theta_1 = \theta_2$)}. \end{array} \right. \end{equation*}
\label{sec:Johnsongraphs}
\end{Definition}
Johnson graphs have been thoroughly studied (\cite{BCN}, \cite{Godsil2}, \cite{Godsil}). In particular, the eigenvalues of $A(v,k)$ are 
$ k(v-k)-j(v-j+1)$ with $j = 0, ..., k$ with degeneracy 
$
\left( \begin{smallmatrix}
v \\ j
\end{smallmatrix} \right)
- \left( \begin{smallmatrix}
v \\ j-1
\end{smallmatrix} \right)
$ \cite{Godsil}. Some pathologies occur when $v \le 2k-1$, as some of the degeneracies become negative or zero. We will see that in our cases, $v$ will always be larger than $2k-1$.
\\ \medskip
For $N$ odd, we write in full generality the states of the second kind as
\begin{equation}
| w \rangle = \prod_{i \in I^w} a_i a_{N-i} \prod_{j_1 \in J_+^w} a_{j_1} \prod_{j_2 \in J_-^w} a_{N-j_2} \sum_{k \in K^w} a_k a_{N-k} |0 \rangle = \sum_{k \in K^w} |w_k \rangle.
\label{eq:secondkind}
\end{equation}

In the previous formula, $I^w$ is the set of integers $i$ in the interval $1,..., (N-1)/2$ such that $w$ contains both the $a_i$ and the $a_{N-i}$ excitation.
 $J_+^w$ ($J_-^w$) is the set of integers $j_1$ ($j_2$), also in the interval $1,..., (N-1)/2$, such that the $a_{j_1}$ ($a_{N-j_2}$) excitation is present but the $a_{N-j_1}$ ($a_{j_2}$) is not (in fact, the sets $J^\pm_w$ are just the sets $K^\pm$ in definition \ref{sec:enslambimpair}). The sets $I^w$, $J_+^w$ and $J_-^w$ are all disjoint. Finally, the sum over $k$, in (\ref{eq:vects2}), was over all integers in $1,..., (N-1)/2$, but since the square of any of the $a$s is zero, the sum really is on $K^w = \{1,..., (N-1)/2\} \setminus (I^w \cup J_+^w \cup J_-^w )$.
 We also define $L^w =  \{1,..., (N-1)/2\} \setminus ( J_+^w \cup J_-^w )$.

Not all states (\ref{eq:secondkind}) are non zero. In fact, because $\langle w_k| w_{k'}\rangle = \delta_{k,k'}$, $\langle w | w \rangle = | K^w |$. If $I^w \cup J^w_+ \cup J_-^w = \{1, ..., (N-1)/2\}$, $K^w$ is empty and the state is zero, as can be seen more easily on (\ref{eq:vects2}).  Recall that we are interested in states that have at most $(N-1)/2$ excitations for $N$ odd, as the number of excitations corresponds to the number of bubbles in the link states, $n$. This imposes that $2 |I^w| + |J_+^w| + |J_-^w| + 2 = n \le (N-1)/2$ (or equivalently, $|L| - 2|I| \ge 2$) and $I^w \cup J_+^w \cup J_-^w \neq \{1, ..., (N-1)/2\}$. 

Two states $|w_1\rangle$ and $|w_2\rangle$ are orthogonal unless $J_{\pm}^{w_1} = J_{\pm}^{w_2}$. We can restrict the study of independence to sets of states with $J_{\pm}^{w_1} = J_{\pm}^{w_2} \equiv J_\pm$ and $|I^{w_1}| = |I^{w_2}| \equiv |I|$ (and so $L^{w_1} = L^{w_2} \equiv L$). In such a set, the states differ only by their $I^{w_1}$ and $I^{w_2}$, and the set has dimension $\left(\begin{smallmatrix} |L| \\ |I|\end{smallmatrix}\right)$. The scalar product of two states $w_1$ and $w_2$ belonging to this set is
\begin{equation}
\langle w_1| w_2\rangle = \left\{ \begin{array}{l l l} |K^{w_1}| & \textrm{if} \quad w_1 = w_2, \\
1 & \textrm{if} \quad |I^{w_1} \cap I^{w_2}| = |I|-1, \\
0 & \textrm{otherwise.}  \end{array}\right.
\end{equation}
The matrix $M(L,|I|)$ of this scalar product is simply $M(L,|I|) = |K| id + A(|L|,|I|)$: $| w \rangle$, with $I^w = \{i_1, i_2, ..., i_{|I|}\}$, is represented by a subset of length $|I|$ of $L$ and is identified to a vertex  of the Johnson graph $J(|L|,|I|)$. The eigenvalues are given by
\begin{equation*}
\underbrace{|L|-|I|}_{|K|} + |I|(|L|-|I|) -j(|L|+1-j), \qquad j = 0, ..., |I|.
\end{equation*}

Because $-j(|L|+1-j)$ is a strictly decreasing function of $j$ on the interval $[0,  |I|]$, the extrema are on the boundaries; they are $ (1+|I|)(|L|-|I|)$ and $|L| - 2 |I|$, both positive. Also because $|L| - 2 |I| >1$, every degeneracy is positive. As all the eigenvalues are positive, there are no null states, and the independence is proved.

For $N$ even, the proof requires a few subtleties. $(S^{+(2)})'$ has a $b_0b_{-1}$ contribution which can be ignored. We therefore consider vectors like (\ref{eq:secondkind}), but with $I^w \cup J_+^w \cup J_-^w\cup K^w = 1, ..., (N-2)/2$, and the possibility to have the $a_{-1}$ excitation. $L^w$ is then defined as $\{1, ..., (N-2)/2 \} \setminus (J_+^w \cup J_+^w)$. The sets of states with and without this excitation, say $S_1$ and $S_2$, can be treated separately because, for any $w_1 \in S_1$ and $w_2 \in S_2$, $\langle w_1| w_2\rangle=0$. For $S_1$, $|L| - 2|I| \ge 1$, and for $S_2$, $|L| - 2|I| \ge 2$. In both cases, all eigenvalues are positive.
\medskip\\
The case $d=0$ is particular. States of the second kind in $S_2$ are
\begin{equation}
| w \rangle =\prod_{i} a_{i}  \sum_{k \in K^w} a_k a_{N-k} |0 \rangle, 
\label{eq:S2states}
\end{equation}
and the product on $i$ has $N/2-2$ terms, all in $\{1,..., N-1 \} \setminus \{N/2\}$. Their number is $\left( \begin{smallmatrix} N-2 \\ N/2-2 \end{smallmatrix} \right)$. These are removed from the states
\begin{equation}| w \rangle =\prod_{i'} a_{i'} |0 \rangle, \label{eq:S2states2}\end{equation}
where the product on $i'$ has $N/2$ terms, also in $\{1,..., N-1 \} \setminus \{N/2\}$. Their number is $\left( \begin{smallmatrix} N-2 \\ N/2 \end{smallmatrix} \right)$. But these two numbers are equal and all states from (\ref{eq:S2states}) are independent from the previous argument. In other words, all the states (\ref{eq:S2states2}) are removed, leaving no degeneracy in $\rho(\mathcal H_N)$. This is the result of proposition \ref{sec:degpair}.
\hfill$\square$\\ 

%
 
\end{document}